\documentclass[sn-mathphys]{sn-jnl}

\newcommand{\de}{\mathrm{d}}

\jyear{2023}%

\usepackage{comment}
\usepackage{float}

\theoremstyle{thmstyleone}%
\theoremstyle{thmstyletwo}%

\theoremstyle{thmstylethree}%

\DeclareMathOperator*{\argmin}{arg\,min}
\DeclareMathOperator*{\argmax}{arg\,max}

\newcommand{\E}{\mathrm{E}}
\newcommand{\Var}{\mathrm{Var}}

\newcommand{\Skew}{\mathrm{Skew}}

\begin{document}

\title[A real world test of Portfolio Optimization with Quantum Annealing]{A real world test of Portfolio Optimization with Quantum Annealing\\
}

\author*[1]{\fnm{Wolfgang} \sur{Sakuler}}\email{wolfgang.sakuler@rbinternational.com}

\author[2]{\fnm{Johannes M.} \sur{Oberreuter}}

\author[3]{\fnm{Riccardo} \sur{Aiolfi}}

\author[3]{\fnm{Luca} \sur{Asproni}}

\author[1]{\fnm{Branislav} \sur{Roman}}

\author[1]{\fnm{Jürgen} \sur{Schiefer}}

\affil[1]{\orgname{Raiffeisen Bank International AG}, \orgaddress{\street{Am Stadtpark 9}, \city{Vienna}, \postcode{1030}, \country{Austria}}}

\affil[2]{\orgdiv{Machine Learning Reply GmbH}, \orgname{Reply SE}, \orgaddress{\street{Luise-Ullrich-Str. 14}, \city{Munich}, \postcode{80636}, \country{Germany}}}

\affil[3]{\orgname{Data Reply S.r.l.}, \orgaddress{\street{Corso Francia 110}, \city{Turin}, \postcode{10143}, \country{Italy}}}


\abstract{In this note, we describe an experiment on portfolio optimization using the Quadratic Unconstrained Binary Optimization (QUBO) formulation. The dataset we use is taken from a real-world problem for which a classical solution is currently deployed and used in production. In this work, carried out in a 
collaboration between the Raiffeisen Bank International (RBI) and Reply, we derive a QUBO formulation, which we solve using various methods: two D-Wave hybrid solvers, that combine the employment of a quantum annealer together with classical methods, and a purely classical algorithm. 
Particular focus is given to the implementation of the constraint that requires the resulting portfolio's variance to be below a specified threshold, whose representation in an Ising model is not straightforward. We find satisfactory results, consistent with the global optimum obtained by the exact classical strategy. However, since the tuning of QUBO parameters is crucial for the optimization, we investigate a hybrid method that allows for automatic tuning.}

\maketitle

\section{Introduction}\label{sec1}


Portfolio Optimization (PO) is a standard problem in the financial industry \cite{markowitz_1952}. 
A monetary budget needs to be completely invested on a given set of financial assets with known historical returns and volatilities. The whole investment amount must be equal to $100\%$ of the initial budget and needs to be split, possibly in different percentages, among the assets. The aim is to maximize the expected return of the resulting portfolio while keeping the risk profile, 
which is measured by the volatility computed from the covariance matrix of the assets,
below a specified limit. Additional constraints may be added to help the diversification of the portfolio over multiple sectors or asset classes.
A similar problem has been solved by \cite{grant_2021}.

The computational complexity of combinatorial optimization problems tend to increase exponentially with the number of variables - here, the number of assets - which at large scale can make solvers incapable of providing only optimal solutions. Instead, the results are likely suboptimal. Currently, it is being investigated in various circumstances whether quantum computers can help cope with this complexity.
In particular, a strategy called \emph{quantum annealing} has proven to be a particularly useful approach to optimization problems \cite{farhi_2000, morita_2008}.

The data used in our work consists of a portfolio structured into three main asset classes: equity (\emph{EQ}), fixed-income (\emph{FI}) and money market (\emph{MM}).
A client portfolio typically ranges from 9 to 11 assets. 
We have chosen this type of dataset because:
\begin{enumerate}
    \item it represents a setup that is actually used in a real-world bank's production environment
    \item it makes it possible to run the optimization in a short amount of time on quantum computers.
\end{enumerate}

Various constraints have to be imposed on the composition of the portfolio, which we describe in greater detail in section \ref{sec:problemformulation}. When imposing constraints in a Quadratic Unconstrained Binary Optimization (QUBO) formulation \cite{grant_2021, lucas_2014, glover_2018}, which is by definition unconstrained, each of these terms must be properly weighted in the objective function such that the resulting solution not only satisfies the constraints, but also maximizes the returns.

We structure this paper as follows: in section \ref{sec:problemformulation} we explain the structure of the problem in detail, we describe the mathematical formulation used by classical algorithms and we introduce the QUBO approach. In section \ref{sec:methodology}, we dive deeper into how we cast our problem as a QUBO and put our work in the context of current research. The results of the various approaches that we use to solve the QUBO are presented in  section \ref{sec:results}, where the different solutions are compared and benchmarked against the exact global optimum.

\section{Problem Formulation}

\label{sec:problemformulation}


We consider the Markowitz portfolio optimization as a quadratic programming problem \cite{markowitz_1952} that determines the fraction $\omega_{i}$ of available budget $B$ to be allocated on the purchase of the $i^{\rm th}$ asset out of potentially $N$ assets with the goal of maximizing returns, while keeping the risk below a target volatility $\sigma_{\rm target}^2$. For simplicity we set $B=1$ and we consider weights $\omega_{i}$ as normalized weights.

The optimization problem is formulated as
\begin{equation}
\label{eq:return}
    \max_{\omega} \{r^T \! \cdot \omega\}
\end{equation}
subject to
\begin{align}
    \omega^{T} \, \Sigma \, \omega & \leq \sigma^{2}_{\rm target} && & \text{(Volatility constraint)} \label{eq:target_vol}\\
    1^{T} \! \cdot \omega & = 1 \, , \, \omega_{i} \geq 0 \, , & \quad \forall i&=1, \ldots, N  &\text{(Weights constraint)}   \label{eq:weights}\\
     A \cdot \omega 
     \, & \langle {\rm op} \rangle \, b \, , & \, \langle \rm op \rangle \, & \in \{ =, \leq, \geq \}
     & \text{(Linear constraints)} \label{eq:lin_const}
\end{align}
where
\begin{itemize}
    \item $r$  is the vector of (mean historical) asset returns
    \item $\omega$  is the vector of asset weights
    \item $\Sigma$ is the covariance matrix of the returns
    \item $\sigma^{2}_{\rm target}$ is the target volatility, i.e.~the maximum allowed risk
    \item $A$ is a matrix of coefficients specifying further linear constraints
    \item $b$ is a vector of constants
\end{itemize}

\subsection{Classical Formulation}


The PO problem investigated in this work is based on a calculation
that is performed in production at Raiffeisen Bank International AG (RBI) as a service for RBI clients. The main objective of the PO is to maximize the expected return while fulfilling several constraints. 
The risk constraint limiting the portfolio volatility is written, using the notation from equation \eqref{eq:target_vol}, as follows: 
\begin{equation}\label{eq:volaconstraint}
\omega^{T} \, \Sigma \, \omega = 
\sum_{i=1}^{N} \sum_{j=1}^{N} \sigma_{ij} \, \omega_{i} \, \omega_{j} \,
\leq \, \sigma^{2}_{\rm target} \, .
\end{equation}
The corresponding risk term is hence a quadratic form that is bounded from above.

We impose several additional linear constraints, either defined globally or for a specific client:
\begin{enumerate}
\item{Normalization constraint: sum of weights $\omega_{i}$ is normalized to 1, i.e. all budget needs to be invested:
\begin{equation}
\sum_{i=1}^{N} \omega_{i} = 1 \, .
\end{equation}
}
\item{Single asset constraints: defining lower and upper bound of the weight of a single asset:
\begin{eqnarray}
\omega_{i} & \geq & \omega_{i, \rm{min}} \, , 
\label{eq:singleassetmin}\\
\omega_{i} & \leq & \omega_{i, \rm{max}} \, .
\label{eq:singleassetmax}
\end{eqnarray}
}
\item{Multi asset constraints: conditions involving a set of assets, e.g.~constraints for a specific asset-class group (EQ, FI or IR):
\begin{equation}
\sum_{i = 1}^{N} a_{j,i} \, \omega_{i} \,\, \langle {\rm op} \rangle \,\, b_{j}, \quad j \in \{ 1, ..., M \}, \quad \langle {\rm op} \rangle \, \in \{ =, \leq, \geq \} \, ,
\end{equation}
}
where $a_{j,i}$ are the elements of matrix $A$, $b_{j}$ are the constants of the constraints $j$, and $M$ is the number of multi asset constraints.
\end{enumerate}

In the classical calculation the strategic asset allocation is accomplished via Markowitz optimization \cite{markowitz_1952},
and the tactical asset allocation is based on the so-called Black-Litterman model \cite{Black7,10.2307/4479577}. 
The Black-Litterman approach takes (i) market expectations derived from market data, 
and (ii) the objective and independent forecasts provided by the bank's internal research group (Raiff\-eisen Research)
as input parameters, and then produces the posterior asset returns and covariance through an optimization process. 
The outputs from the Black-Litterman process are then used as input for the strategic Markowitz optimization.

For the classical calculations the statistical software programming environment R is employed.
The optimization algorithm is run via IBM's CPLEX optimization software package,
which provides solvers for linear and quadratic programming problems \cite{cplex_2022}. 
These CPLEX solvers can be called from the R environment via the R interface module ``Rcplex'' \cite{rcplex_2022}.

Since the mathematical optimization model represents a convex problem, and furthermore the volume of the data sets currently faced in RBI's production environment is rather small, the classical optimization procedure is able to quickly find the exact solution, which is the global optimum.
Therefore, it clearly cannot be the goal of the study to achieve a more accurate result.
This solution rather serves as ultimate target goal to be ideally achieved by the optimization procedure executed on a quantum computer. 
The study serves as a starting point to apply the working quantum algorithms to improve results, where classical solutions are not satisfactory. 

\subsection{QUBO Formulation}\label{QUBO}

The Quadratic Unconstrained Binary Optimization (QUBO) model represents a wide range of combinatorial optimization problems \cite{lucas_2014, grant_2021, glover_2018}. It is currently the most applied model in the quantum computing area for these kind of problems. The QUBO model is expressed by the following optimization problem:
\begin{equation}
    \text{minimize} \, f_{Q}(x) \, ,
\end{equation}
where $f_{Q}: \{0,1\}^{n} \xrightarrow{} \mathbb{R}$,
\begin{equation}
    f_{Q}(x) = \sum_{i=1}^{n} \sum_{j=i}^{n} q_{ij} \, x_{i}x_{j} \, ,
\end{equation}
is a quadratic polynomial over binary variables $x_{i} \in \{0,1\}$ and coefficients $q_{ij} \in \mathbb{R}$ for $1 \leq i \leq j \leq n$. The QUBO problem consists of finding a binary vector $x^{*}$ that is minimal with respect to $f$ among all other binary vectors, namely
\begin{equation}
  x^{*} = \argmin_{x \in \{0,1\}^{n}} f_{Q}(x) \, .
\end{equation}
In order to maximize $f_{Q}(x)$, one simply minimizes $f_{-Q}(x) = - f_{Q}$. 

Another, more compact way to formulate $f_{Q}(x)$ is using matrix notation,
\begin{equation}
     f_{Q}(x) = x^{T} Q \, x \, ,
\end{equation}
where $Q \in \mathbb{R}^{n \times n}$ is a square matrix  containing the coefficients $q_{ij}$. It is common to assume an upper triangular form for $Q$ since it is a symmetric matrix, thus the transformation can always be achieved without loss of generality with simple tricks. Many problems can be effectively re-formulated as a QUBO model by introducing quadratic penalties into the objective function as an alternative to explicitly imposing constraints in the classical sense \cite{glover_2018}. The penalties introduced are chosen so that the influence of the constraints on the solution process can alternatively be achieved by the natural functioning of the optimizer as it looks for solutions that avoid incurring the penalties. For a minimization problem, these penalties are used to create an augmented objective function to be minimized.

\section{Methodology}\label{methodology}

\label{sec:methodology}
In this work we tackle the PO problem by modeling it as a QUBO
\cite{grant_2021, mugel_2020, mugel_2022}. 
This formulation enables the use of special-purpose quantum computers, quantum annealers, to find the minimum of a given objective function. 

In recent years, along with the developments in the quantum computing field, increasing attention has been drawn to the formulation of well-known combinatorial optimization problems as a QUBO model or, equivalently, Ising model \cite{lucas_2014, glover_2018}. The equivalence consists in the solution of one of the two models also being the solution of the second one, up to a linear change of variables: this allows to adhere to common formulations in operations research that exploit binary variables taking values in $\{0,1\}$, while being able to exploit quantum annealers to find the minimum of the optimization problem at hand. 
Therefore, a current focal point in the quantum optimization literature is to examine the capabilities of quantum annealers in application to combinatorial optimization problems \cite{PhysRevE.58.5355,2001Sci...292..472F,10.1007/s10878-014-9734-0,PhysRevX.5.031026,doi:10.1126/science.aaa4170,2017arXiv170206248H,Asproni2020}. 

The PO problem is a key activity in the financial services industry \cite{mugel_2022}. The classical Markowitz model is a convex quadratic programming problem, which in its simplest form, the Mean-Variance model, has a polynomial worst-case complexity bound \cite{nemirovski_1994, kerenidis_2019}, where the algorithm's running time $t$ behaves as:
\begin{equation}
t \sim O(N^k), \quad 2 \leq k \leq 4 \, . 
\end{equation}
 Numerical calculations using state-of-the-art classical optimization algorithms indicate that the classical Markowitz model shows at best a quadratic time complexity with respect to the number of assets \cite{brown_2008, pedersen_2021}. However, the complexity of enhanced PO problems, e.g.~the so-called Limited Asset Markowitz (LAM) model (also called cardinality constrained Markowitz model), depends on the specific constraints that are additionally imposed on the basic objective \cite{maringer_2008, cesarone_2009, cesarone_2011}. Additional constraints increase the level of complexity, which can result at worst in an NP-hard problem whose complexity scales exponentially as the number of assets grows \cite{bienstock_1995, jin_2016}:
\begin{equation}
t \sim O({\rm e}^{N}) \, . 
\end{equation}

 This, combined with the nonlinear nature of the problem that particularly fits the QUBO formulation, has led to the use of a quantum computing approach to tackle the problem \cite{Rosenberg_2016}. In this work we build a QUBO model similarly to \cite{grant_2021}. We follow the approach by including a risk measure constraint on the assets' covariances, given by \eqref{eq:target_vol} 
and include the left-hand side term of the inequality in the QUBO formulation, fine-tuning the model parameters such that the overall risk does not exceed $\sigma_{\rm target}^2$ (cf. \ref{subsubsec:targetvolatility} for details).
Finally, we discretize the continuous variables $\omega$ into a set of binary variables, each of which is weighted in the QUBO by a coefficient. Differently from the approach proposed in \cite{grant_2021}, for each asset $i$ and the corresponding variable $\omega_i$, our discretization uses a fixed number of binary variables, representing a given interval $[\omega_{i, {\rm min}}, \omega_{i, {\rm max}}]$ which may differ from asset to asset. This entails the possibility to use a reduced number of variables to represent assets' weights in the portfolio, while on the other hand potentially providing a different granularity for different assets. Further mathematical details are explained in Section~\ref{sec:disc}.

\subsection{Discretization of Variables}\label{sec:disc}

In order to cast the problem into the QUBO formulation, one needs to choose a binary encoding of the weights. As the weights are fractions, the exponents in the binary expansion of the weights are going to be negative. Using discrete rather than continuous variables inevitably limits the accuracy of the solution. While the accuracy increases if more binary variables are being used for each weight, so do the resource needs. Thus, one has to carefully find an optimal number of binary variables that represents a trade-off between target accuracy and acceptable resource usage. 

If we allow $K$ variables for the discretization of each weight $\omega_i$, $i=1,\ldots, N$, the upper bound of the number of QUBO variables for the discretization without considering any additional (slack) variable is $N \! \cdot \! K$. 

However, the number of variables needed can be reduced after carefully analysing the linear constraints that restrict the weights for single assets, the so-called single-min and single-max constraints, see \eqref{eq:lin_const}, \eqref{eq:singleassetmin} and \eqref{eq:singleassetmax} of Section~\ref{sec:problemformulation}, respectively. For example, if the weight of an asset is limited within a specific range, fewer binary variables are needed to achieve the same granularity covering only that range
\begin{equation}
    \omega_i = \omega_{i, {\rm min}} + (\omega_{i, {\rm max}} - \omega_{i, {\rm min}}) \cdot \omega'_i \, ,
    \label{eq:disc_weight}
\end{equation}
where the normalized weight $\omega'_i$ is restricted to 
\begin{equation}
0 \leq \omega^{\prime}_{i} \leq 1 \, .
\end{equation}
With the definition
\begin{equation}
    \Delta \omega_i = \omega_{i, {\rm max}} - \omega_{i, {\rm min}} \, ,
\end{equation}
one gets
\begin{equation}
    \omega_i = \omega_{i, {\rm min}} + \Delta \omega_i \cdot \omega'_i \, .
    \label{eq:omega_i}
\end{equation}
In a binary expansion using $K$ bits, i.e. having a granularity $p_{K} = 1/2^{K}$, $\omega^{\prime}_{i}$ is given as
\begin{equation}
\label{eq:z}
    \omega^{\prime}_{i} = \sum_{k=1}^{K} 2^{k-1}x_{i,k} \, p_{K} \, ,
\end{equation}
where $x_{i,k} \in \{0,1\}$, $i = 1, \ldots, N$, $k = 1, \ldots, K$, are binary variables.

Naturally, the granularity would be chosen to be $p_{K} = 1/2^{K}$. Then, the normalized weight $\omega'_i$ in \eqref{eq:disc_weight} is effectively restricted to
\begin{equation}
0 \leq \omega^{\prime}_{i} \leq (1 - p_{K}) \, ,
\end{equation}
and the effective granularity $p_{K \! , {\rm eff} \, i}$ is given by
\begin{equation}
p_{K \! , {\rm eff} \, i} = \Delta \omega_i \cdot p_{K} \, .
\end{equation}
While this choice for the granularity technically does not allow to reach exactly $\omega_{\rm max}$ for each asset, we can reach a number close to it by summing up all the terms, incidentally ensuring by design to have $\omega_i < \omega_{i,\rm max}$ such that the max-constraint is automatically fulfilled. An alternative choice of $\tilde{p}_{K} = 1/2^{K-1}$ for the granularity would not have these advantages but would allow to reach the maximum amount exactly. This would also come at the cost of using one binary variable more for each asset.

For example, if $K=10$, the granularity of the normalized weight $\omega^{\prime}_{i}$ is $p_{K } = 1/2^{10}$, and the effective granularity for the weight $\omega_{i}$ of asset $i$ is $1/2^{10} \cdot \Delta \omega_i = 9.765625 \cdot 10^{-4} \, \cdot \Delta \omega_i$ of the budget. The maximum fraction of the normalized weight is $\sum_{i=1}^{K} 2^{i-1-K} = 1 - 1/2^{K} \approx 0.999023$, thus giving for asset $i$ an effective maximum weight
\begin{equation}
\omega_{i, {\rm max\_eff}} = \omega_{i, {\rm max}} - 9.765625 \cdot 10^{-4} \, \cdot \Delta \omega_i \, .
\end{equation}
However, if we use $p_{K'} = 1/2^{20}$, the effective granularity for asset $i$ is approx. $9.5367 \cdot 10^{-7} \, \cdot \Delta \omega_i$ of the budget, while the maximum fraction is $1 - 1/2^{K'} \approx 0.999999046$.

The choice of $K$ relies on the effective granularity needed, the level of approximation manageable and the number of variables implementable.

Finally, we use the same number of variables for all assets, although the discretized ranges vary among different assets, since the effective granularity $p_{K \! , {\rm eff} \, i}$ for asset $i$ is given by $p_K \cdot \Delta \omega_i$. 
That means that the effective granularity for each asset is in fact finer than $p_K$, because in our real-world setup $\Delta \omega_i$ is always smaller than 1 for all assets (in the current setup one has $\omega_{i, {\rm max}} < 1$, and $\Delta \omega_i = 0.1$ for all $i$, thus the effective granularity is the same for all assets, i.e. $p_{K \! , {\rm eff} \, i} = p_{K \! , {\rm eff}}$).
A possible improvement when having the same granularity for each asset would be to reduce the number of binary variables for each asset.

The error in representing the individual weights due to the finite granularity also leads to an error in the total budget invested. This is because every weight is in principle a random rational number between $0$ and $1$. When approximating the $\omega_i'$ in any weight in a binary expansion, this number will be represented with an error depending on $p_K$.  
Treating these errors as a distribution, we are calculating the expected value and variance of the error in the Appendix~\ref{sec:constraintviolation}. The resulting expectation value of the error $\epsilon$ is 
\begin{equation}
\E[\epsilon] = \frac{p_K^2}2 \, ,
\end{equation}
and standard deviation is 
\begin{equation}
\Var[\epsilon] = \frac{p_K^2}{12} + \frac{p_K^3}4 - \frac{p_K^4}{4} \;.
\end{equation}

However, given our construction, we are sampling the interval between the minimal and maximally allowed values, only as explained in \eqref{eq:omega_i}. Therefore the error accumulates to 
\begin{equation}
    \delta \omega = \delta \sum_i \omega_i = \sum_i \Delta \omega_i \, \delta \omega_i' = \delta \omega' \sum_i \Delta \omega_i \;,
    \label{eq:minmaxerror}
\end{equation}
where in the last step we have used that fact that all the $\omega_i$ are constructed in the same way according to \eqref{eq:z}. This makes the error dependent on the individual min-max constraint, more precisely on the difference between maximum and minimum. 
%

Keeping this in mind is important for the interpretation of the results of our experiments in Section~\ref{sec:results}.

\subsection{Structure of Objective Function}
\label{subsec:objectivefunction}
Considering the terms including the constraints mentioned above, the objective function consists in our case of the following four terms:
\begin{itemize}
    \item Returns to be maximized, ref. eq. \eqref{eq:return}
    \item Weights constrained as all budget needs to be invested, ref. eq. \eqref{eq:weights}
    \item Linear constraints, ref. eq. \eqref{eq:lin_const}
    \item Target volatility constraint, ref. eq. \eqref{eq:target_vol}
\end{itemize}
This formulation allows to consider a single QUBO expression made up of 4 terms:
\begin{equation}\label{eq-qubo}
    f_Q = \lambda_{1} H_{1} + \lambda_{2} H_{2} + \lambda_{3} H_{3} + \lambda_{4} H_{4},
\end{equation}
where $\lambda_{l} > 0$ is the penalty coefficient incorporating the relative importance of the $l^{\rm th}$ term and the sign linked to the maximization or minimization; the $H_{l}$ is the Hamiltonian derived from the QUBO matrix  of the $l^{\rm th}$ term.
We analyze each term in detail below.

\subsubsection*{Returns $H_{1}$}
The optimization of returns consists of minimizing
\begin{equation}
    H_{1} = - r^{T} \omega,
\end{equation}
where $r= (r_{1},\ldots, r_{N})$ is the vector of returns of assets $i=1, \ldots, N$ and $\omega= (\omega_{1}, \ldots, \omega_{N})$ is the vector of asset weights. The "return" term $H_{1}$ is the basic objective term of the optimization problem. In order to formulate the problem in the QUBO framework one needs to discretize the weights as described in Section \ref{sec:disc}. Without losing generality the discretization expressed in \eqref{eq:disc_weight} is used.
With this discretization the QUBO formulation of the basic objective term $H_{1}$ is written as \begin{equation}
    H_{1} = - \sum_{i=1}^{N} \left( \omega_{i, {\rm min}} + (\omega_{i, {\rm max}} - \omega_{i, {\rm min}}) \cdot \sum_{k=1}^{K} p_K 2^{k-1} x_{i,k}\right) \cdot r_{i}.
\end{equation}

\subsubsection*{Weights constraint $H_{2}$}
This term is a hard constraint on the sum of investments of the initial budget and it is expressed as:
\begin{equation}
    1^{T} \! \cdot \omega = \sum_{i=1}^{N} \omega_{i} =  1.
\end{equation}
With the discretization explained in Section \ref{sec:disc} the QUBO formulation becomes
\begin{equation}
    H_{2} = \left[\sum_{i=1}^{N} \left( \omega_{i, {\rm min}} + (\omega_{i, {\rm max}} - \omega_{i, {\rm min}}) \cdot \sum_{k=1}^{K} p_K 2^{k-1}x_{i,k}\right) - 1\right]^{2} .
\end{equation}

\subsubsection*{Linear constraints $H_{3}$}
This term represents a set of linear constraints defined by the matrix $A$ and the vector $b$ in \eqref{eq:lin_const}. These constraints are slightly different from $H_{2}$ because they include inequalities. One can 
always put them into a QUBO formulation by including auxiliary variables, so-called slack variables, which are also represented as a binary expansion using slack binary variables. Supposing that matrix $A$ is of the type $M \times N$ where $M$ is the total number of linear constraints and $N$ the number of assets, the QUBO formulation becomes
\begin{equation}
\label{eq:linear}
    \sum_{j=1}^{M} \lambda_{3j} \left (\sum_{i=1}^{N} a_{j,i} \, \omega_{i} + 
    \alpha_{j} \, s_{j} - b_{j}\right)^{2},
\end{equation}
where $a_{j,i}$ are the elements of matrix $A$, $b_{j}$ are the constants of the various linear constraints $j$, $s_{j}$ are slack terms, that are introduced to transform inequality constraints effectively into equality conditions, and 
$\alpha_{j}$ are the signs related to the slack terms, that depend on the relational operators of the constraints:
\begin{equation}
    \alpha_{j}=\begin{cases}
1\qquad & \text{if $j^{\rm th}$ constraint is }\leq\\
0\qquad & \text{if $j^{\rm th}$ constraint is }=\\
-1\qquad & \text{if $j^{\rm th}$ constraint is }\geq
\end{cases},
\end{equation} 
Using a binary formulation the slack term $s_j$ is given as
\begin{equation}
\label{eq:linear_slack}
 s_{j} = \beta_{j} \, \sum_{k=1}^{S_{j}} p_{S_{j}} 2^{k-1} s_{j,k} \, ,
\end{equation}
where $s_{j,k} \in \{0,1\}$, $j = 1, \ldots, M$, $k = 1, \ldots, S_{j}$, are the actual binary slack variables, 
and $\beta_{j}$ is the maximum value of the continuous version of the slack term, given by
\begin{equation}
    \beta_{j}=\begin{cases}
\argmax\limits_{x_i \in \{0,1\}} \,\, [ \, (b_{j} - \sum_{i} a_{j,i} \, x_{i}), 0 \, ]  \qquad & \text{if $j^{\rm th}$ constraint is }\leq\\
\argmax\limits_{x_i \in \{0,1\}} \,\, [ \, (\sum_{i} a_{j,i} \, x_{i} - b_{j}), 0 \, ]  \qquad & \text{if $j^{\rm th}$ constraint is }\geq
\end{cases}.
\end{equation} 
The number of slack variables $S_{j}$ for constraint $j$ depends on the effective slack term granularity 
\begin{equation}
p_{S_{j}, {\rm eff}} = \beta_{j} \cdot p_{S_{j}} \, , \quad p_{S_j} = 1/2^{S_{j}} \, .
\end{equation}
For practical reasons the number of slack variables $S_{j}$ has been fixed for each linear constraint $j$ to the number $K$ of physical binary variables per asset. Thus, while the number of slack variables is always the same for the various constraints, the effective slack term granularity $p_{S_{j}, {\rm eff}}$ varies.  


Another approach which has been performed in the current activity is related to those linear constraints that act individually on each asset: these give lower and upper bounds on the values of such assets. In this way, the discretization described in equation \eqref{eq:disc_weight} can be applied to the new range defined by the constraints. This allows to satisfy those constraints by construction and there is no need to include them in the QUBO formulation, leading to an ease of calibration and findings of feasible solutions. For example, if two constraints impose that the investment of an asset $i$ must be within the range $[\omega_{i, {\rm min}}, \omega_{i, {\rm max}}]$, then the discretization will find a fraction of the value $(\omega_{i, {\rm max}} - \omega_{i, {\rm min}})$.

With the discretization explained in Section \ref{sec:disc} eq.~\eqref{eq:linear} is transformed into
\begin{align}
\begin{split}
    \sum_{j=1}^{M} \lambda_{3j} \left [\sum_{i=1}^{N} \left( a_{j,i} \, \omega_{i, {\rm min}} + a_{j,i} \, (\omega_{i, {\rm max}} - \omega_{i, {\rm min}}) \cdot  \sum_{k=1}^{K} p_K 2^{k-1}x_{i,k}\right) + \right.  \\
      \left. + \, \alpha_{j} 
 \, \beta_{j} \sum_{k=1}^{S_{j}} p_{S_{j}} 2^{k-1}s_{j,k} - b_{j} \right]^{2} .
      \end{split}
\end{align}

\subsubsection*{Target volatility constraint $H_{4}$}
\label{subsubsec:targetvolatility}
The QUBO formulation of this term strictly depends on the approach implemented to consider the target volatility constraint as in Section \ref{sec:problemformulation}. 

As a first example we consider the constraint rewritten in eq. \eqref{eq:target_vol} in which the target volatility is set to zero. In this way, the portfolio risk is handled via the minimization of the term
\begin{equation}
    \omega^{T} \, \Sigma \, \omega  = \sum_{i=1}^{N} \sum_{j=1}^{N} \sigma_{ij} \, \omega_{i} \, \omega_{j} \, ,
\end{equation}
which in the QUBO formulation with the discretization explained in Section \ref{sec:disc} becomes
\begin{align}
\begin{split}
    \sum_{i=1}^{N} \sum_{j=1}^{N} \Tilde{\sigma}_{ij} \cdot \left ( \omega_{i, {\rm min}} + (\omega_{i, {\rm max}} - \omega_{i, {\rm min}}) \cdot \sum_{k=1}^{K} p_K 2^{k-1}x_{i,k} \right) \cdot \\ \cdot \left (\omega_{j, {\rm min}} + (\omega_{j, {\rm max}} - \omega_{j, {\rm min}}) \cdot \sum_{k=1}^{K} p_K 2^{k-1}x_{j,k} \right),
    \end{split}
\end{align}
where $\Tilde{\sigma}_{ij}$ in an adjusted coefficient such that
\begin{equation}
    \Tilde{\sigma}_{ij} = 
    \begin{cases}
    \sigma_{ij} \quad & \text{if} \quad i=j\\2\sigma_{ij} \quad & \text{if} \quad i < j \\ 0 \quad & \text{otherwise}
    \end{cases}  
\end{equation}

When the maximal risk is not allowed to exceed a threshold value given by the target volatility $\sigma^{2}_{\rm target}$ the following less-than-or-equal constraint  has to be fulfilled 
\begin{equation}
\label{vol_cons}
\omega^{T} \, \Sigma \, \omega \leq \sigma^{2}_{\rm target}\, .
\end{equation}
Putting this constraint into a non-constraint QUBO-like form one gets:
\begin{equation}
\label{H4}
H_{4} = \left ( \omega^{T} \, \Sigma \, \omega + s_{\rm vola} - \sigma^{2}_{\rm target} \right )^{2} \, ,
\end{equation}
where, since a less-than-or-equal constraint (and not an exact equality condition) has to be handled, a slack variable term $s_{\rm vola}$ has to be introduced which, using binary slack variables $s_{\sigma,k} \in \{0,1\}$, $k = 1, \ldots, S_{\sigma}$, is given by
\begin{equation}
s_{\rm vola} = \sigma^{2}_{\rm target} \, \sum_{k=1}^{S_{\sigma}} p_{S_{\sigma}} \, 2^{k-1} \, s_{\sigma,k} \, ,
\end{equation}
in which $S_{\sigma}$ is the number of binary slack variables for the volatility constraint that depends on the effective volatility slack term granularity \begin{equation}
p_{S_{\sigma}, {\rm eff}} =  \sigma^{2}_{\rm target} \cdot p_{S_{\sigma}} \, , \quad p_{S_{\sigma}} = 1/2^{S_{\sigma}} \, .
\end{equation}

After squaring (\ref{H4}) one obtains
\begin{equation}
H_{4} = (\omega^{T} \, \Sigma \, \omega)^{2} + 2 \, (s_{\rm vola} - \sigma^{2}_{\rm target}) \, \omega^{T} \, \Sigma \, \omega - 2 \, s_{\rm vola} \, \sigma^{2}_{\rm target} + s^{2}_{\rm vola} + \sigma^{4}_{\rm target} \, .
\end{equation}
Since the volatility risk term is itself a quadratic form the first term contains quartic and cubic contributions in $\omega$.
Strictly speaking the problem is no longer a QUBO (quadratic) problem, but it turned into a so-called PUBO (Polynomial Unconstrained Binary Optimization) problem (sometimes also called HUBO for Higher Order Unconstrained Binary Optimization) \cite{glover_2011a, glover_2011b, palmer_2021}. 
The existence of up to fourth-order polynomial terms represents a fundamental complication in the procedure since these terms cannot be mapped onto the Ising model of the quantum computer, whose interactions are by definition restricted to linear 1-body and quadratic 2-body terms. 
However, several workarounds for the PUBO problem are proposed in the literature:
\begin{enumerate}
    \item {Applying Order-Reduction techniques \cite{mugel_2020}: using this method is rather expensive since one needs additional bits. \\}
    \item {Linearization \cite{palmer_2021}: in this approach the quadratic volatility term is replaced by a linearized expression.
    \begin{equation}
    H_{4} = \left ( k^{T} \, \Sigma \, \omega - \sigma^{2}_{\rm target} \right )^{2} \, ,
    \end{equation}
    where $k$ is a vector of constants which are called linear weights. Due to the linearization the whole term remains quadratic. However, finding an appropriate value of $k$ is somehow arbitrary: one option is to find $k$ in a self-consistent way, another possibility is to fine-tune $k$ starting from a convenient value like $k_{i} = 1/N \, \forall i$. \\   
    }
    \item{Replacement by a Equality-to-Zero condition \cite{grant_2021}: if the target volatility threshold value is sufficient small, then it can be approximated by zero. When the right-hand-side is exactly zero, then the $\leq$ operator can be replaced by the equality operator since the left-hand-side, the quadratic volatility term, is positive-definite. A constraint like $g(\omega) = 0$ with a positive definite function $g(\omega) \geq 0$ can be handled in the optimization model very easily by just adding a term $\lambda \, g(\omega)$ to the objective. Thus, in this approach one has
    \begin{equation}
    H_{4} = \omega^{T} \, \Sigma \, \omega \, ,
    \end{equation}}
\end{enumerate}

For the calculations employing (i) the classical Qbsolv solver, and (ii) \mbox{D-Wave's} Hybrid BQM solver we used the latter workaround, namely replacement of the PUBO term by an equality-to-zero constraint. By tuning the weight of the volatility constraint carefully, i.e.~by choosing an appropriate Lagrange multiplier, we obtain a feasible formulation while being able to optimize the complete objective function including all the other constraints.

The calculations performed with (iii) {D-Wave's} new Hybrid CQM (Constrained Quadratic Model) solver do not need such a replacement, because the CQM solver can handle both linear and quadratic conditions naturally as genuine constraints.

\subsection{Computational Method}


To solve the PO problem different approaches have been followed up such that the capabilities of quantum and quantum-inspired solutions could be thoroughly assessed and benchmarked. In order to check and quantify the quality of these solutions, the results have been compared with the global minimum of the optimization problem, which, given the limited size of the data at hand, could be easily found via classical strategies.

First and foremost, the QUBO model as described in Section~\ref{methodology} has been built and D-Wave's QBSolv library has been exploited to solve the optimization problem through classical optimization techniques. In order to do so, the Binary Quadratic Model (BQM) data structure has been used that stores each entry of the QUBO model, assigning biases and couplers as penalty coefficients to each variable and pair of variables, respectively.

Second, we have investigated the use of D-Wave's Hybrid Binary Quadratic Model (BQM), which decomposes the overall QUBO problem into 
subproblems suitable to be solved on a Quantum Processing Unit (QPU). Those subproblems can be solved directly on the QPU, thus having the benefit of Quantum effects such as Quantum tunneling to best find high quality solutions. The decomposition step is needed in order to have QUBO subproblems of sufficiently small size that match current QPU architecture; this procedure is handled automatically by D-Wave's Hybrid software.

With the aforementioned QUBO solvers, one crucial step needed to find not only feasible but also optimal solutions, is to fine-tune some significant QUBO parameters, namely the Lagrange multipliers $\lambda_l$ from (\ref{eq-qubo}), that act as relative weights between the various optimization terms and constraints that build up the whole QUBO expression. This step is non-trivial and might lead to suboptimal solutions, especially when the number of optimization terms and constraints, and thus the overall complexity of the problem, increases. 

As next step, and particularly motivated to overcome the problem to fine tune Lagrange parameters, we have investigated the usage of D-Wave's next generation Hybrid solution, the Constrained Quadratic Model (CQM) solver.
The CQM solver is the newest product of D-Waves's hybrid solver family. It enables to formulate constraints in their natural form as 'they are', i.e. as real constraints 
\begin{equation}
\text{term} \, \langle\text{op}\rangle \, b, \quad \langle\text{op}\rangle \, \in (=, \le, \ge) \, .
\end{equation}
where 'term' can be any linear or quadratic form in the binary variables.
Thus, even the maximal volatility constraint (\ref{vol_cons}), which is quadratic and a less-than-or-equal condition (i.e.~not an equality condition), can be entered into the CQM solver directly without any modification.
Before, when using one of the previous solvers, e.g. the Hybrid BQM solver, one had to formulate constraints as penalty terms,
which in case of a inequality condition even had to be supplied with an auxiliary variable term $\alpha \, s$. The constraint had to be put into a parabolic form multiplied by a Lagrange multiplier $\lambda$ (cf. \ref{subsec:objectivefunction}):
\begin{equation}
    \lambda \, (\text{term} + \alpha \, s - b)^2 \, , \quad \alpha = \{0, 1, -1\} \, \, {\rm if} \, \, \langle\text{op}\rangle = \{=, \le, \ge\}  .
\end{equation}

With the availability of the CQM solver from D-Wave, constraints are handled automatically. 
However, since the implementation details of D-Wave's CQM algorithm have not been publicly revealed by the software vendor, from a software end-user's perspective it remains hidden under the surface how the constraints are in fact implemented or formulated.

\section{Results}
\label{sec:results}

We have performed our investigation on multiple solutions ranging from classical strategies adopted by the QBSolv library to hybrid techniques for decomposing the optimization problem into suitable subproblems which are solved both on Classical and Quantum Processing Units (QPUs).
In order to exploit the full potential of the available software for quantum optimization, and thus to reach the highest performing solution strategy, two available strategies have been investigated, namely using a Binary Quadratic Model (BQM) and a Constrained Quadratic Model (CQM).

The former is used as a data structure to represent the QUBO modeling and hence underlies the same principles: one needs to fine tune the model parameters in order to find feasible and optimized solutions, in terms of maximum return and minimum volatility.
The latter allows to explicitly declare which terms of the optimization are genuine constraints and which constitute the objective function. The management of different terms is then delegated to the hybrid solver library, i.e.~to the software side, and thus allows the software user to reduce the time spent calibrating QUBO parameters.

\begin{figure}[ht!]
    \centering
    Volatility vs. Return \\
    \includegraphics[width=0.49\textwidth]{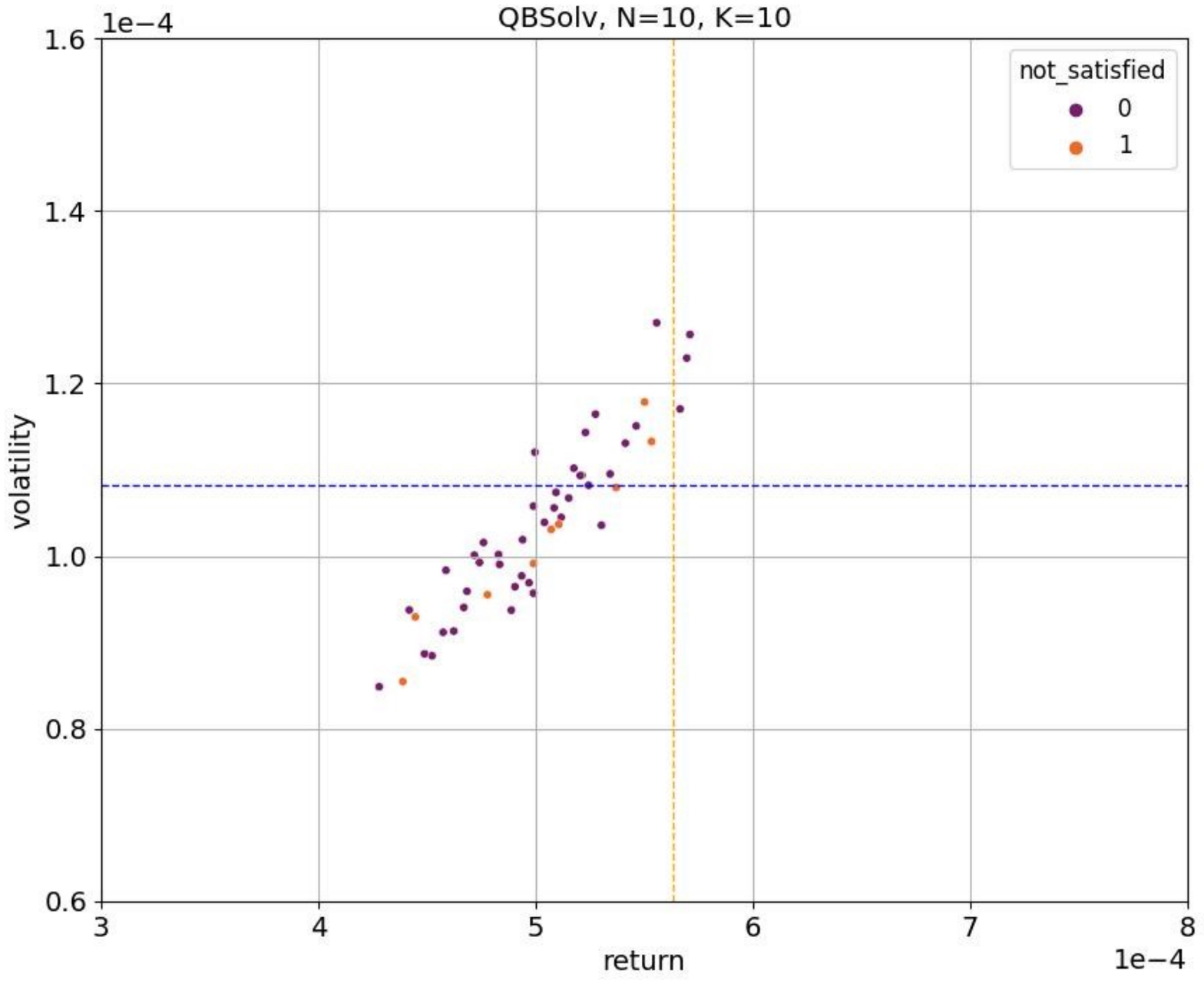}
    \includegraphics[width=0.49\textwidth]{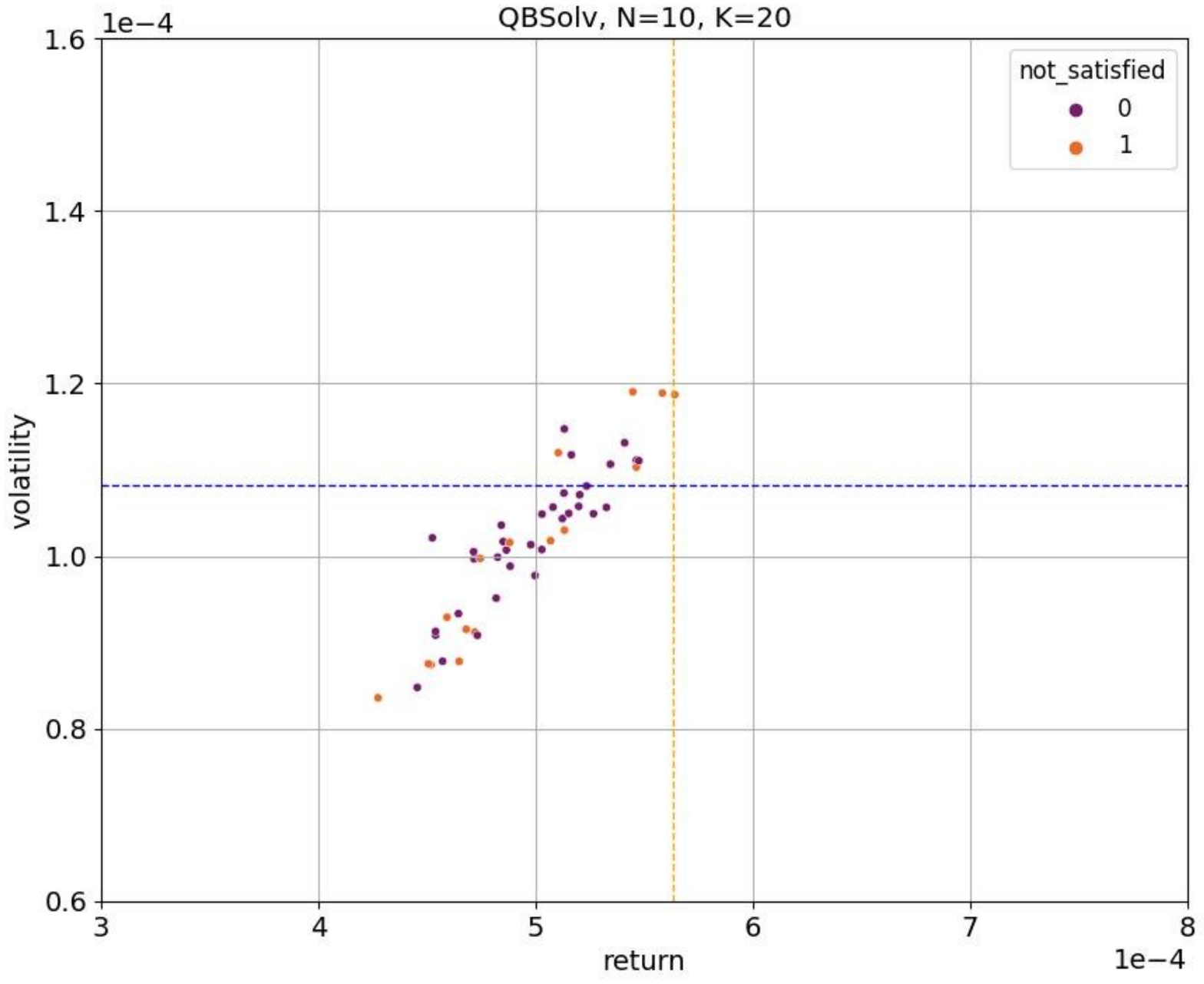} \\
    \includegraphics[width=0.49\textwidth]{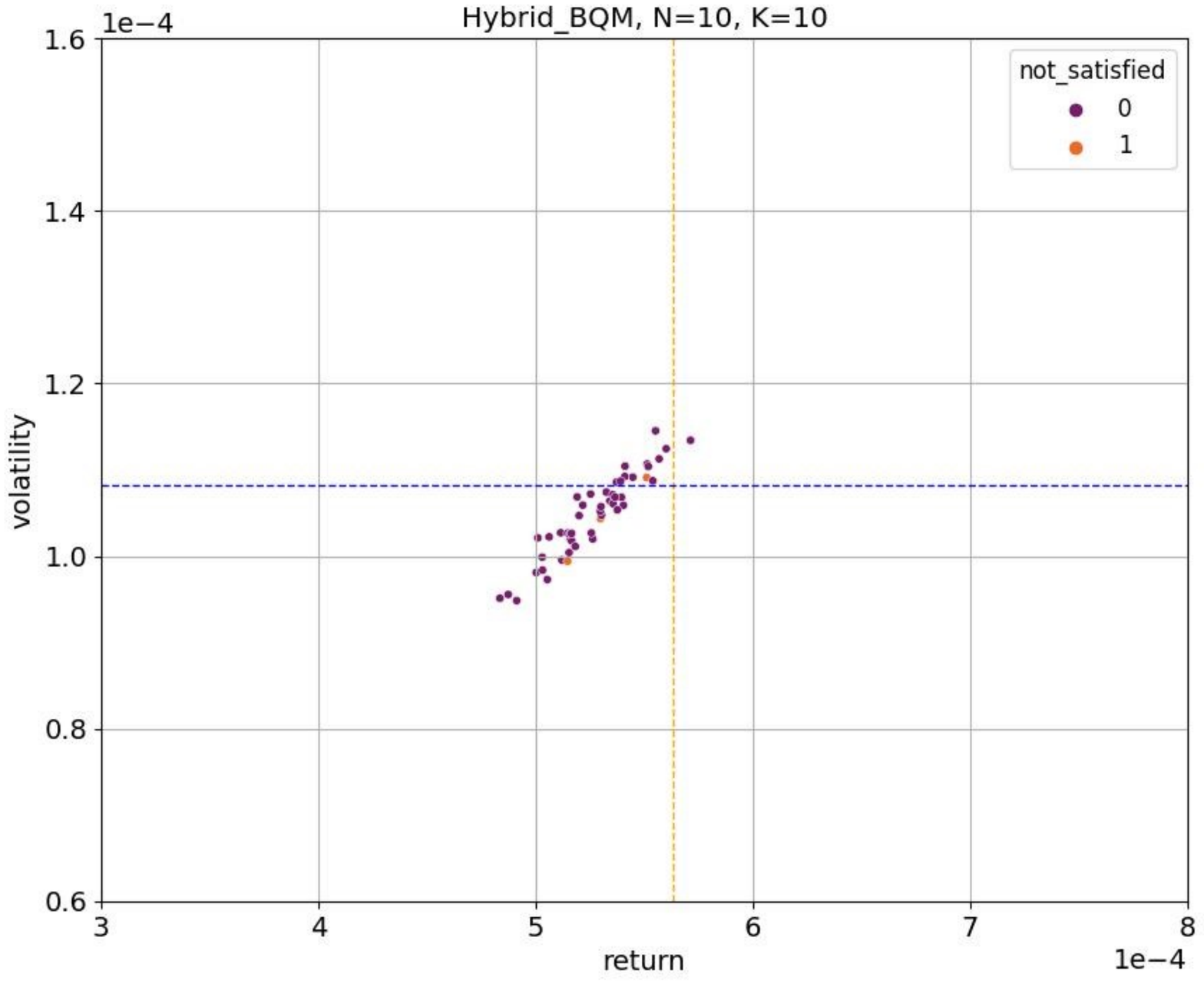}
    \includegraphics[width=0.49\textwidth]{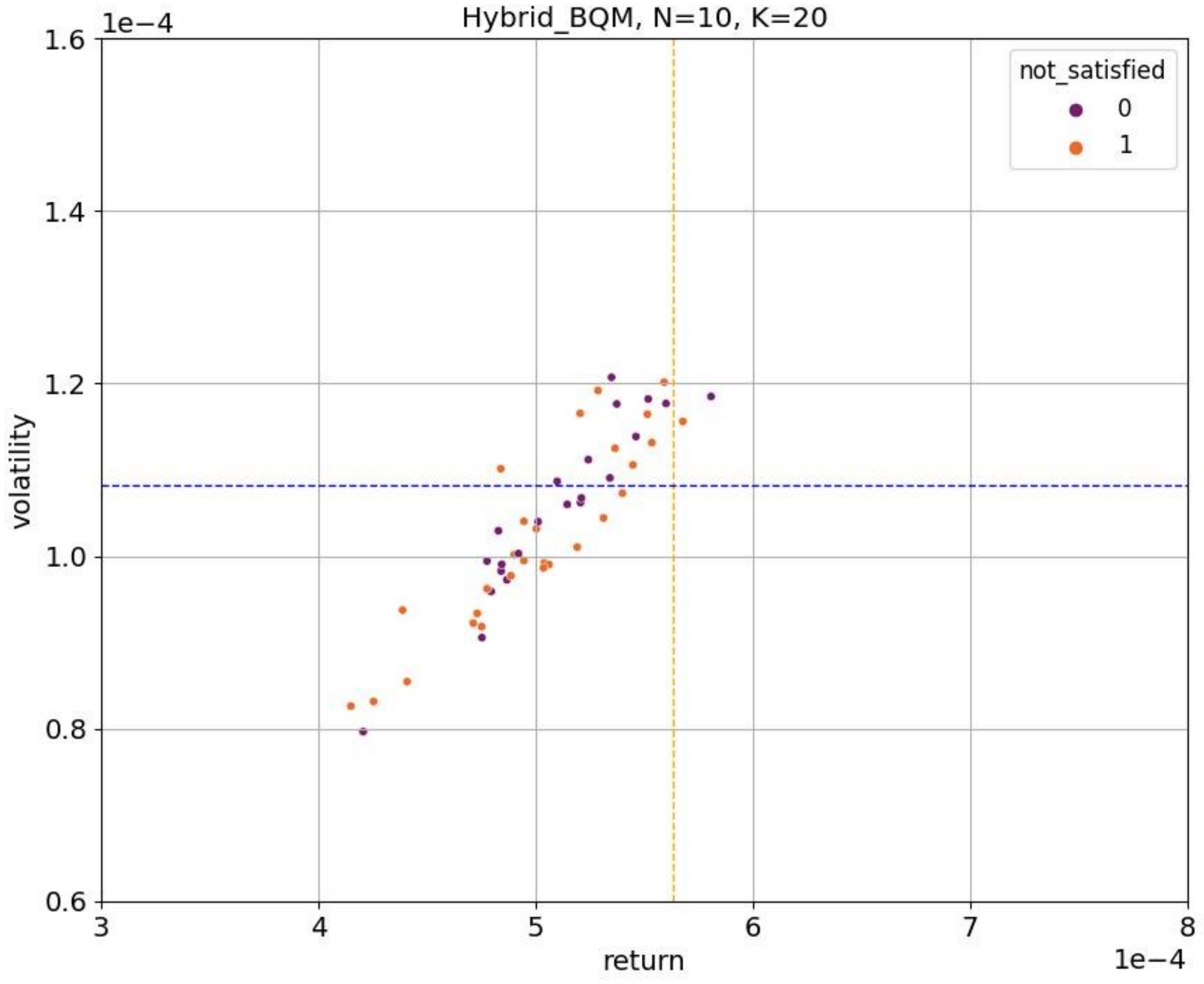} \\
    \includegraphics[width=0.49\textwidth]{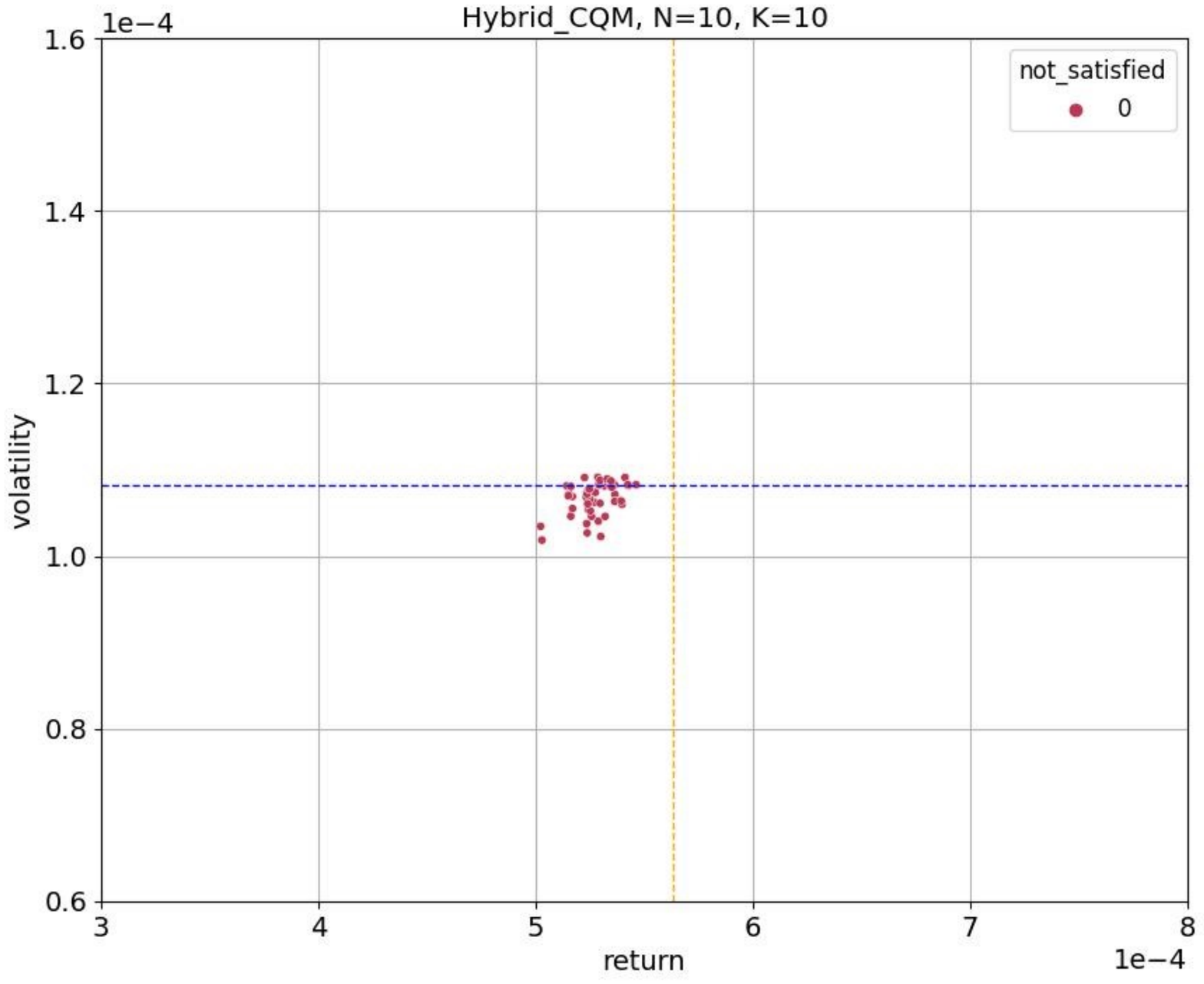}
    \includegraphics[width=0.49\textwidth]{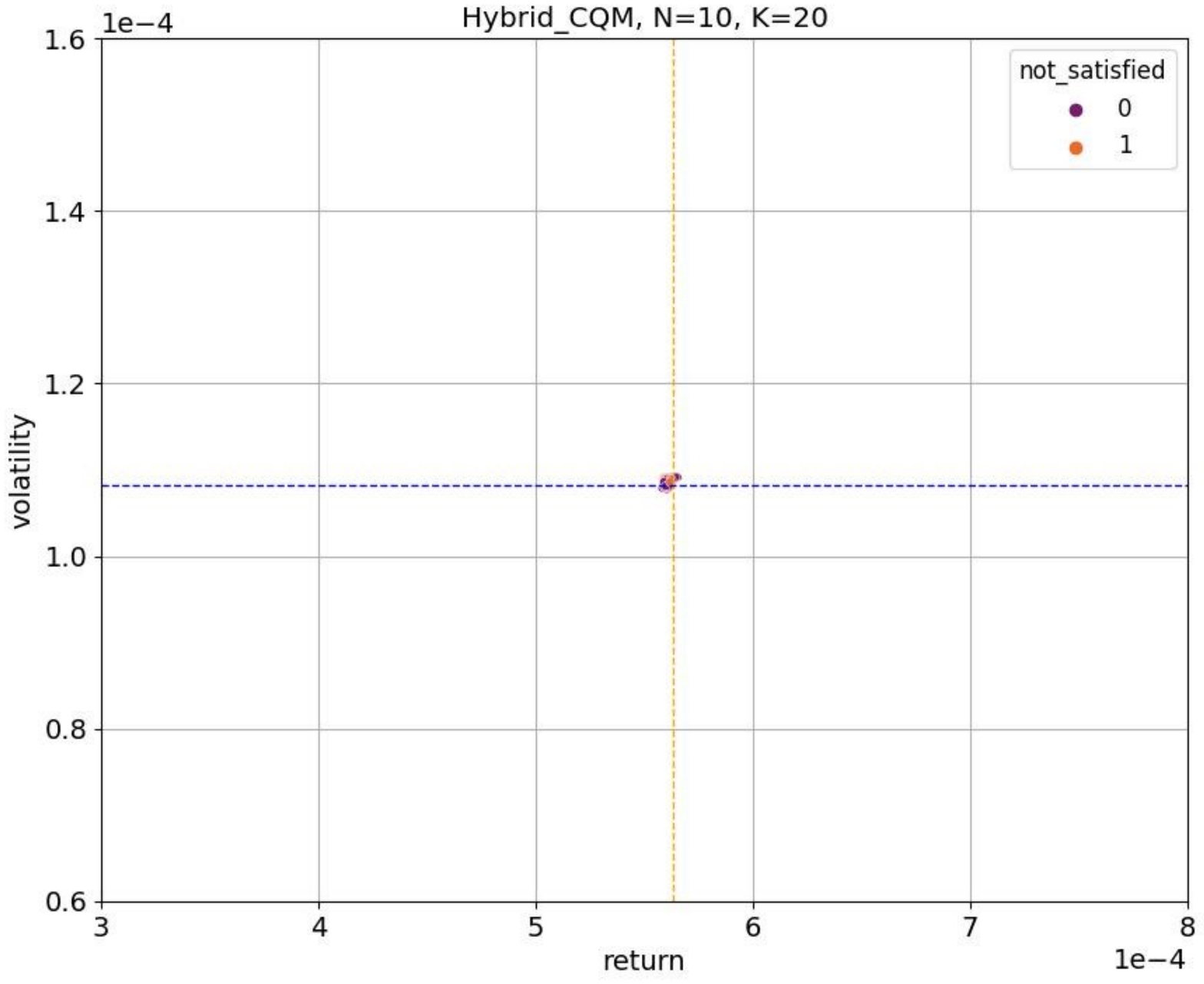}
    \caption{
    Distribution of solutions of several runs for different solvers and granularity. The solvers employed are the purely classical QBSolv, and the Hybrid BQM and Hybrid CQM, that both have a quantum backend; $N$ denotes the number of assets considered; $K$ represents the number of binary variables, i.e.~the number of qubits, used for each asset. This number also  determines the granularity of the calculation. The dashed lines, blue for the volatility, and yellow for the expected return of the portfolio, represent the solutions obtained by the classical CPLEX solver. The $\textit{not\_satisfied}$ label on each solution indicates the number of constraints violated as described in paragraph \ref{sec:disc}. The spread of the solutions along an upwards slope makes sense from a business point of view, insofar as higher risk should be associated with higher return. It can be clearly seen that the results obtained with the Hybrid CQM solver are most precise. A zoomed version of these plots with better visibility is given in Figure \ref{fig:result_scatter_zoomed}.
    }
    \label{fig:result_scatterplot}
\end{figure}

In the following paragraphs we show multiple results using common notation and considerations:

\begin{itemize}
    \item The scatter plots in Figure~\ref{fig:result_scatterplot} show the achieved return vs. volatility for different parameter sets, where one data point represents the best result of one experiment. The dashed vertical yellow line marks the result achieved by the classical solver, which is expected to be close to the theoretical optimum of the return with the given volatility, marked by a dashed blue horizontal line. Points higher than the blue line represent experiments which have yielded impermissible results (risk too high). Results to the right of the yellow line would represent experiments which yield better performing portfolios than the ones classically found, which is not expected. We are looking for best experiment in the lower left quadrant, i.e. the one closest to the intersection of the dashed lines. \\
    \item The $\textit{not\_satisfied}$ label in the plots refers to the number of constraints violated. 
    The investment constraint is considered satisfied in those cases where the actual sum of investments deviates from the target ($100\%$) no more than a small amount which is given by the effective granularity $p_{K, {\rm eff}}$:
    \begin{equation}
    \left\lvert \sum_{i=1}^{N} \omega_{i} - 1 \right\lvert \, \leq \, p_{K, {\rm eff}}
    \end{equation}
    
    The distribution of the sum of approximated weights is shown 
    in Figure~\ref{fig:asset_weights} as measured in our experiments, which describes to what extent the normalization constraint is fulfilled. \\
    \item Our notation is such that:
    \begin{itemize}
        \item $\textit{N}$ refers to the number of assets
        \item $\textit{K}$ refers to the number of (regular) qubits
        \item $\textit{t}$ refers to the maximum time provided to CQM to retrieve the solution
    \end{itemize}
\end{itemize}






\subsection{Solvers comparison on business KPIs}

In this paragraph we focus on the comparison of the results based on the business KPIs, namely volatility and return of the optimized portfolios.
Figure~\ref{fig:result_scatterplot} reports volatility vs return plots (zoomed version with better visibility on details in Figure \ref{fig:result_scatter_zoomed} in the appendix). The classical optimization yields a volatility and return value reported as a horizontal dashed blue line and a vertical dashed yellow line, respectively. The optimal solution lies at the intersection of the two lines. The dots represent the results (samples) from the QUBO formulation. Given the flexible nature of QUBO problems not setting constraints explicitly and given our approach to satisfy the volatility constraint, it is in principle possible to find results that do not satisfy such constraint. These solutions are represented in the plot with the dots lying above the dashed blue line. From a business perspective, these are solutions that must be discarded. We have however included them in the plots to report a detailed and complete overview of the outcome of the QUBO problems with a calibration of the QUBO weights as thorough as possible (note that the calibration has not been implemented when using the CQM solver).
Among the samples found via the different solvers, we were able to find results fairly close to the global optimum found via a classical optimization procedure.
\begin{figure}
    \centering
    \includegraphics[width=\textwidth]{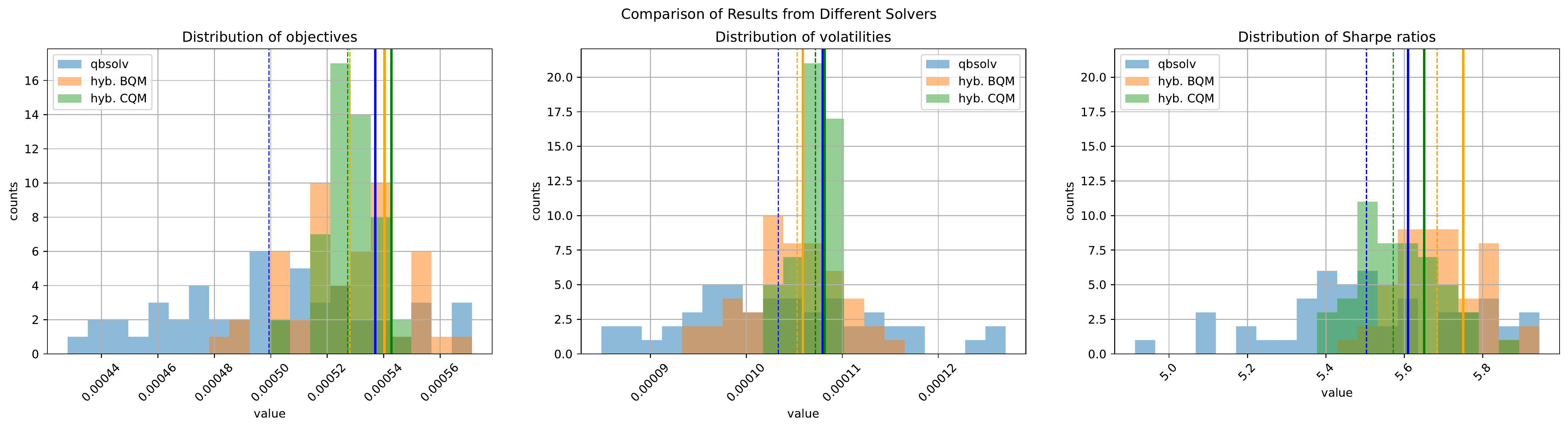}
    \includegraphics[width=\textwidth]{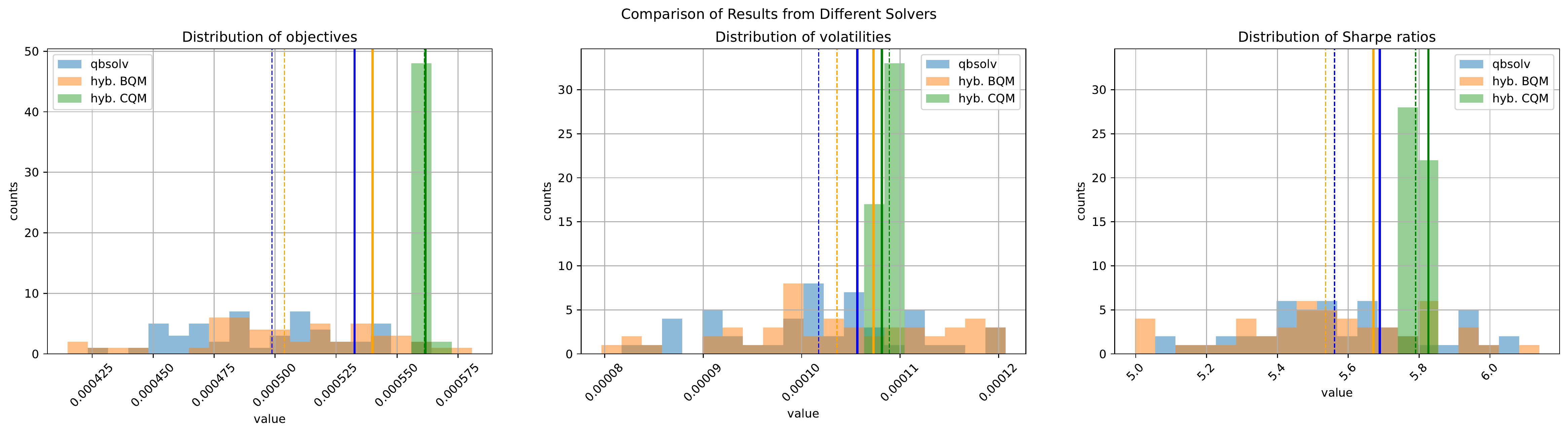}
    \caption{We compare the distribution of business relevant characteristics of portfolios obtained with different solvers. The solid lines of each color mark the classical result. The dotted lines mark the experiment yielding the highest objective while the volatility is below the set threshold, i.e. the best permissible portfolio obtained with these parameters. First row $K=10$, second row $K=20$.}
    \label{fig:allsolvers_w10}
\end{figure}

To produce the bottom plots of Figure~\ref{fig:result_scatterplot} regarding CQM performances, we have excluded the volatility constraint from the counting of the number of constraints not satisfied, which is shown via the dots' label within the plot. This is due to the fundamentally quartic nature of such constraint and the difficulty in treating this term within a QUBO formulation, which requires advanced procedures and cannot be reformulated as a quadratic term without the use of additional variables. 


In Figure~\ref{fig:allsolvers_w10}, we see a comparison of business KPIs for the three solvers considered in this work, namely QBSolv, Hybrid BQM and Hybrid CQM. We see that hybrid solvers perform better than QBSolv. While the average objective is higher for Hybrid CQM, so is the average risk. This has to be put into the perspective that behind the Hybrid BQM and QBSolv solution there is the need to fine tune the QUBO weights, or Lagrange Multipliers, which is handled automatically in the Hybrid CQM. Looking at the Sharpe ratio, we see that Hybrid BQM actually outperforms the other solvers. However the objective was to maximize the returns while satisfying the volatility constraint, rewritten as a risk minimization, and the Sharpe ratio has not been introduced as an explicit term of the objective functions. It also needs to be pointed out that the spread of the values is in the range of $10^{\rm th}$ permille and thus very reduced, almost rendering the approaches on par.

\subsection{CQM capabilities scaling with the number of assets}

\begin{figure}
    \centering
    \includegraphics[width=0.49\textwidth]{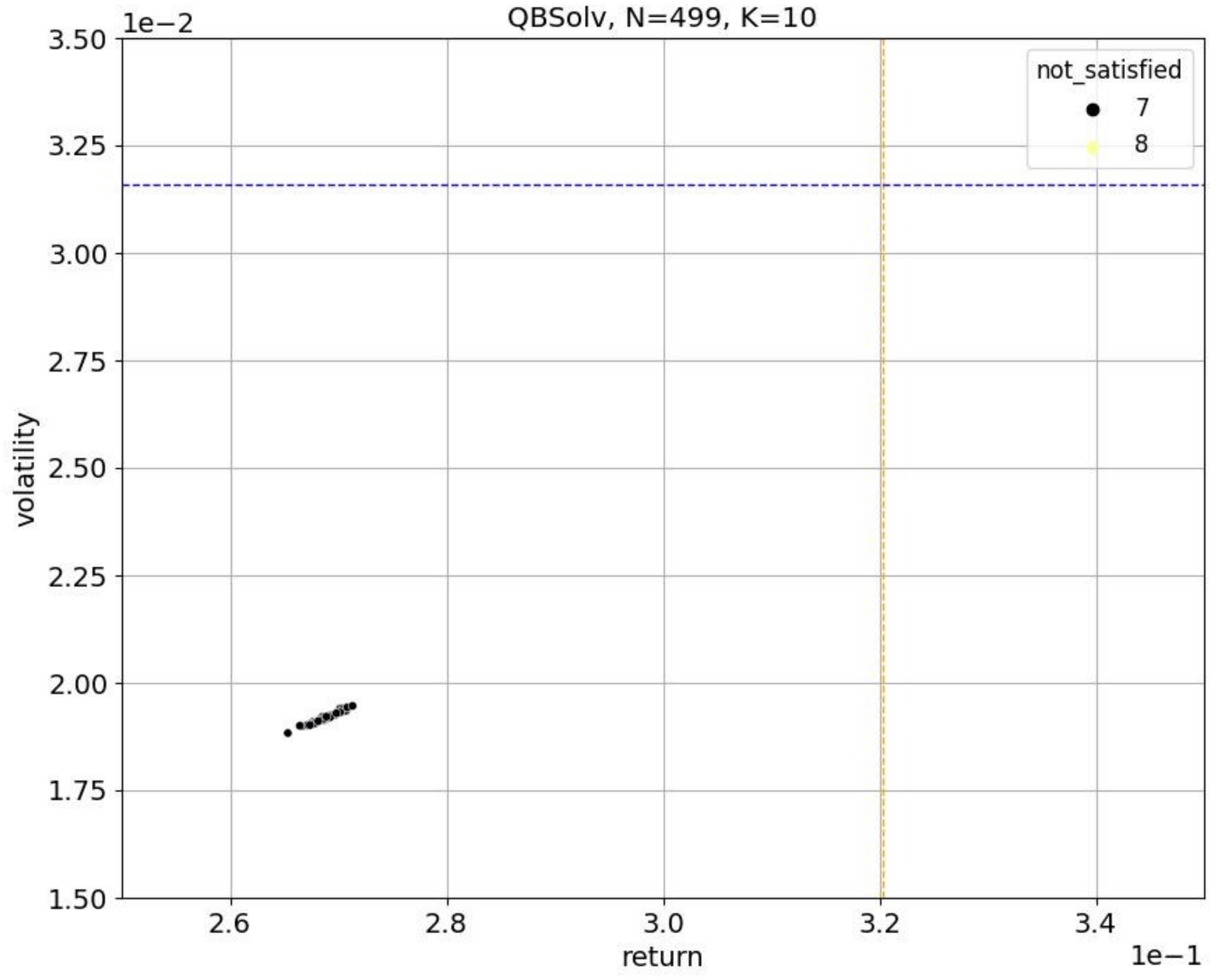}
    \includegraphics[width=0.49\textwidth]{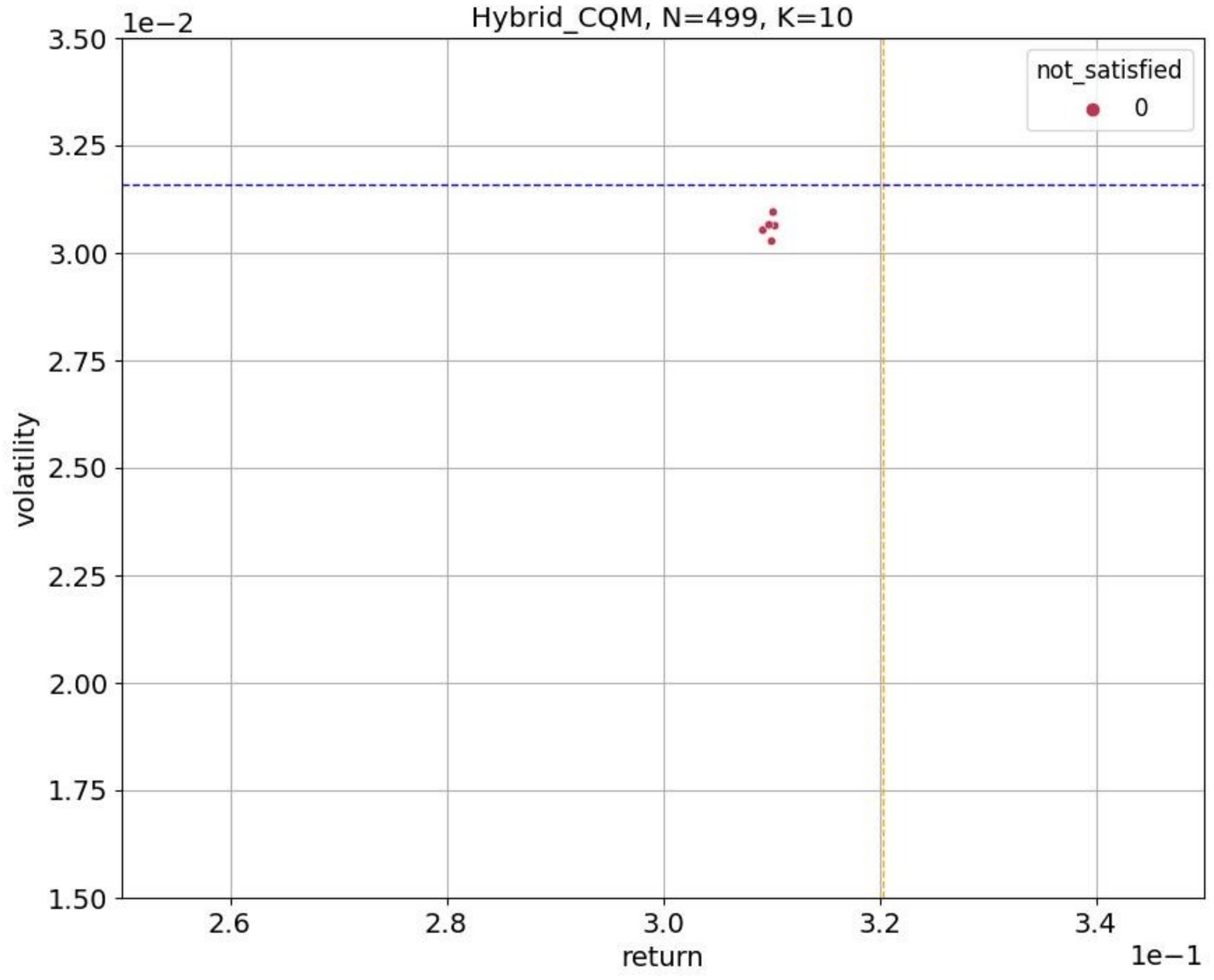}
    \caption{Comparison of risk vs return for a more complicated problem with 499 assets. This problem could only be solved with QBSolv (left) and hybrid CQM (right).}
    \label{fig:scatter_assetscaling}
\end{figure}

\begin{figure}
    \centering
    \includegraphics[width=\textwidth]{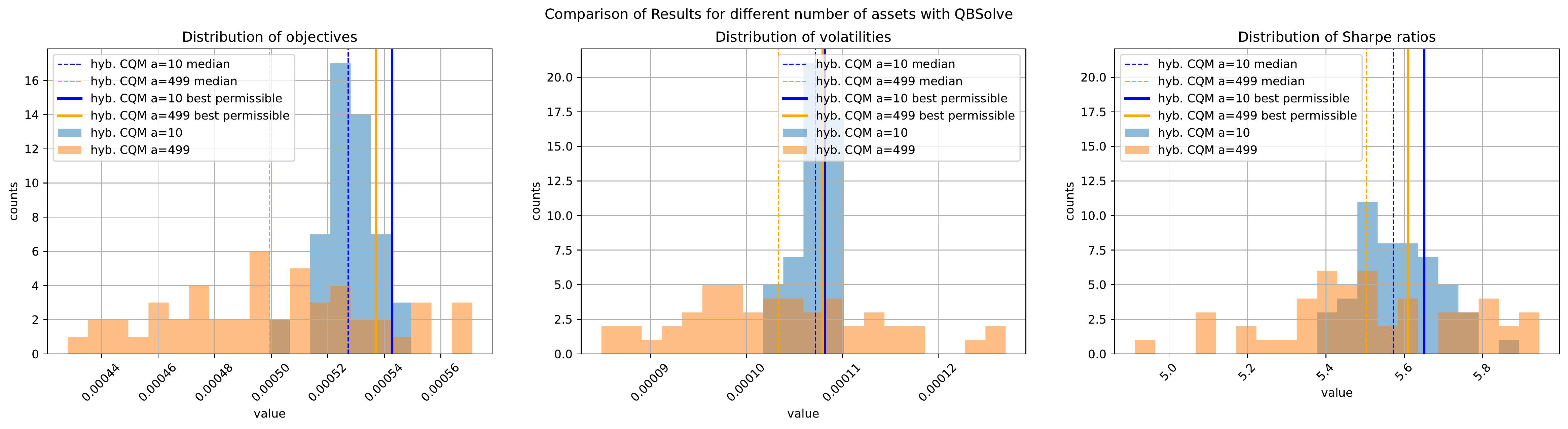}
    \caption{Comparison of performance of Hybrid CQM scaled by the number of assets. Performances are shown as the portfolios' expected returns, volatilities and Sharpe Ratios, respectively.}
    \label{fig:assetscaling}
\end{figure}

Scaling up with the number of assets is crucial for industrial applications and thus we investigate the capabilities of the considered solvers both for $10$ and $499$ assets in the initial pool of assets.

Figure~\ref{fig:scatter_assetscaling} shows the comparison of results in terms of returns and volatility of QBSolv and CQM approaches, benchmarked against the classical solution, for $499$ assets. Not only is QBSolv likely to find infeasible solutions, but they are also lower quality with respect to CQM results: the CQM's feature to automatically handle constraints proves to be a consistent approach to obtain feasible portfolios. At the same time, the objective (i.e. the expected return) is also close to the classical benchmark.

Figure~\ref{fig:assetscaling} shows the distribution of expected return (objective), volatility and the derived Sharpe Ratios for multiple runs of the optimization via QBSolv. For each, we report both the distributions (histograms) related to $10$ and $499$ asset and the median as well as the best result (in terms of objective, volatility and Sharpe Ratio, respectively) of a feasible portfolio.
The results we find are twofold:
\begin{enumerate}
    \item The best results that we find are close to the classical solutions;
    \item The distribution of solutions is more peaked for $10$ assets, while for $499$ assets it takes a more flattened shape across multiple values.
\end{enumerate}

In particular, the second result is expected: as we scale up with the number of assets, so does the number of variables in the QUBO and thus the complexity of the problem. We show that the CQM solver is able to find high quality solutions, whilst needing to run multiple times before finding the portfolio that optimizes our measures.
From a business perspective the best solution would be the one that maximizes the expected return (while satisfying the volatility constraint). Alternatively, even though not directly optimized, the best portfolio would be the one that maximizes the Sharpe Ratio.

\subsection{Solvers capabilities scaling with the number of qubits}

\begin{figure}
    \centering
    \includegraphics[width=0.32\textwidth]{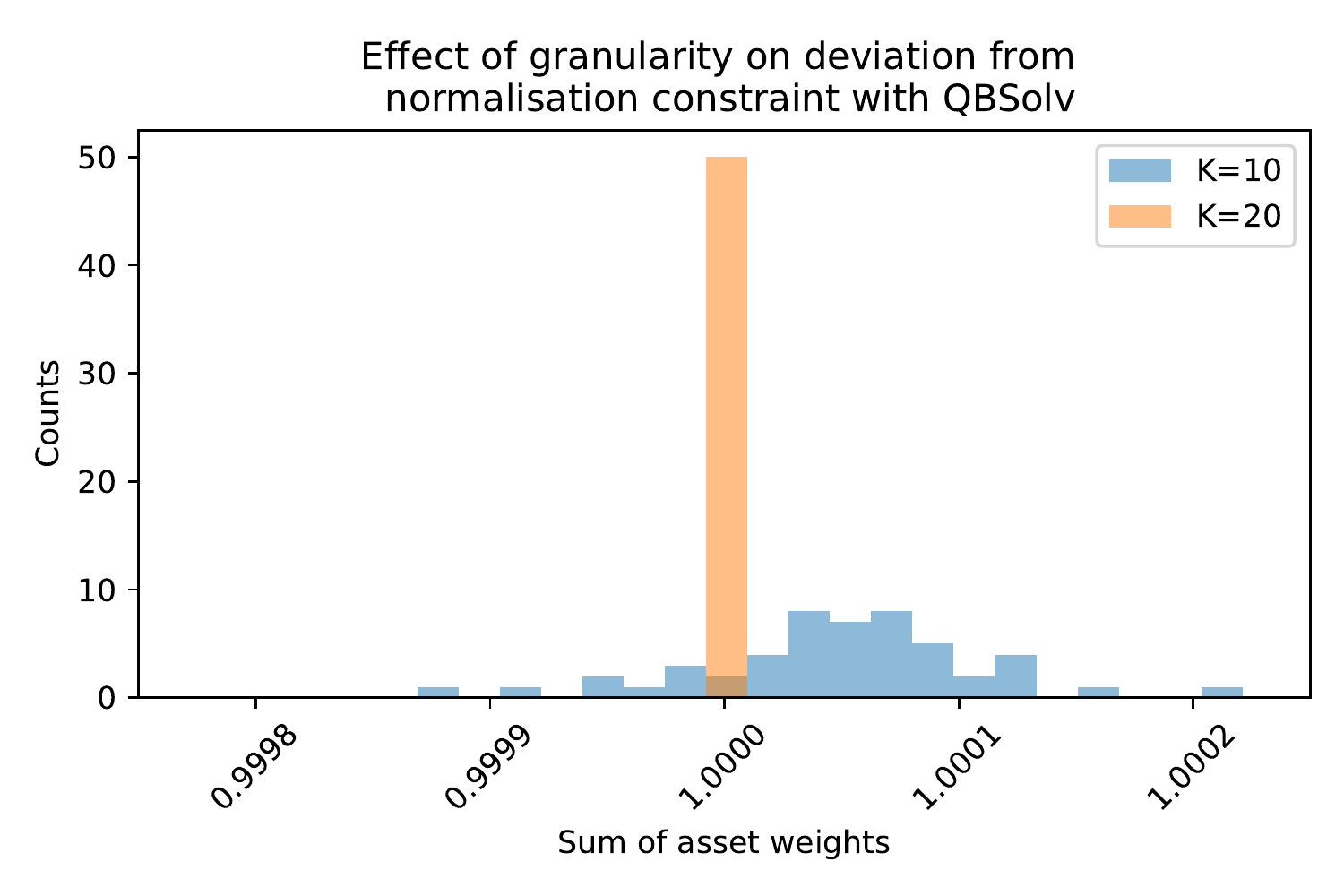}
    \includegraphics[width=0.32\textwidth]{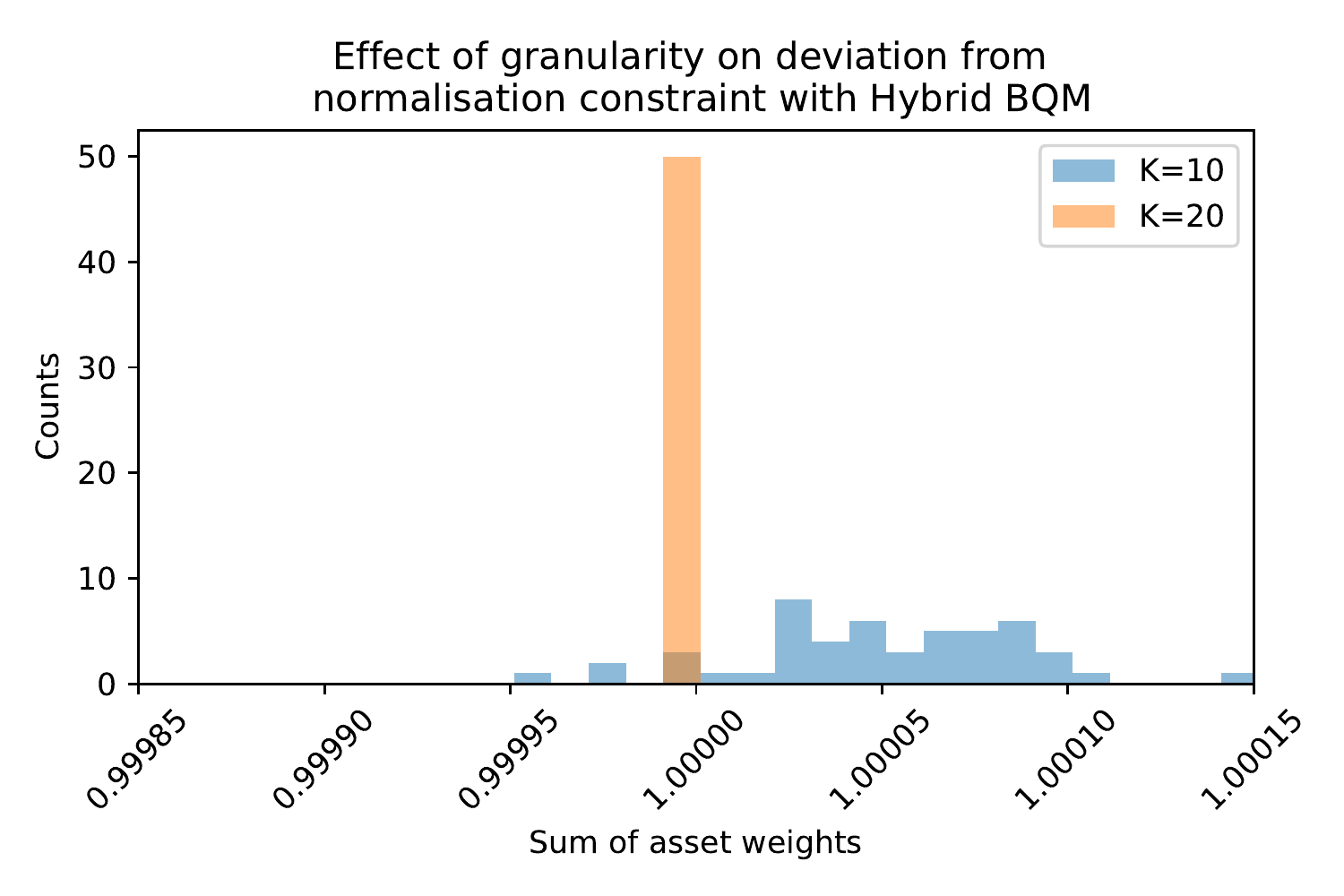}
    \includegraphics[width=0.32\textwidth]{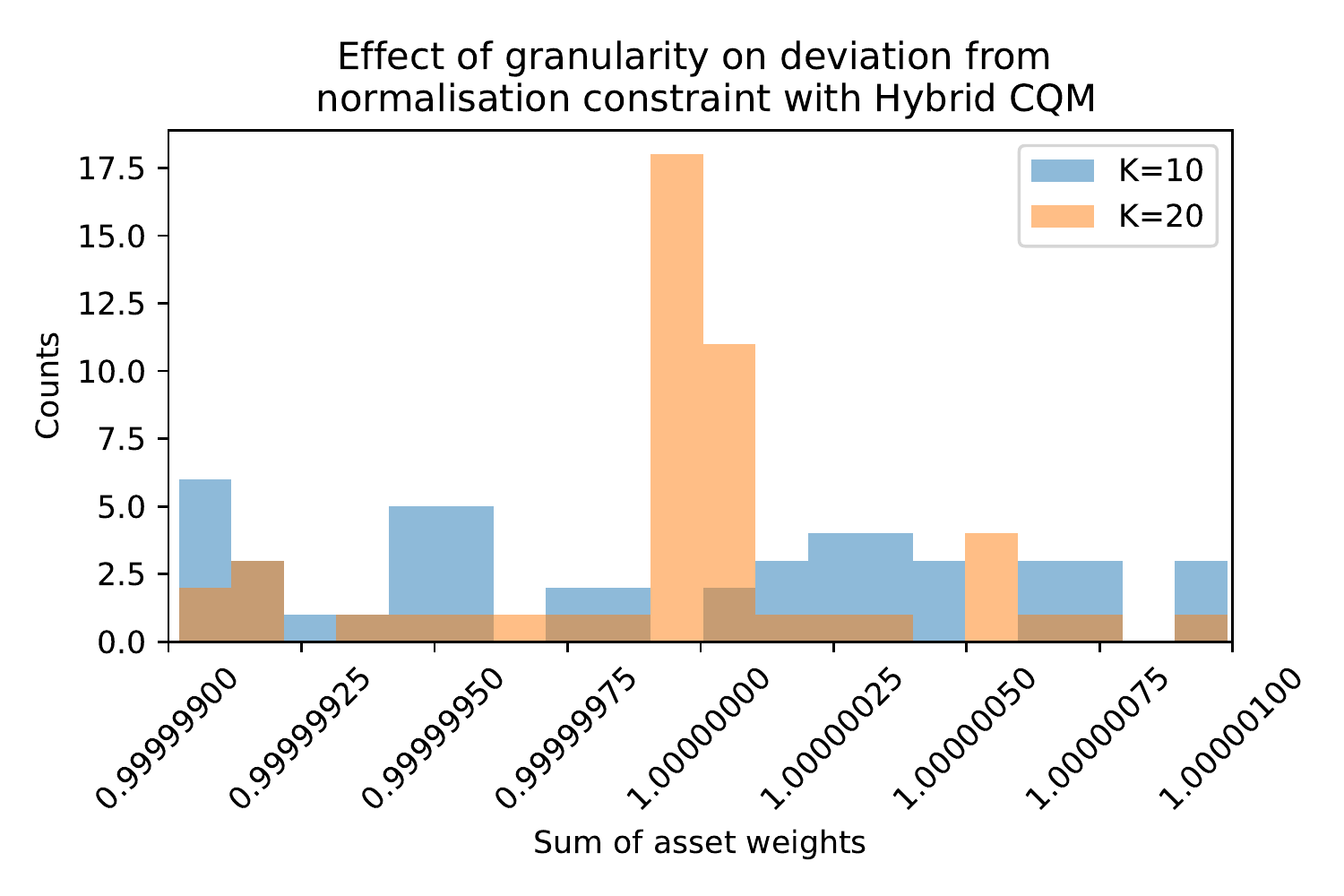}
    \caption{Comparison of the distribution of portfolio weights using $K=10$ and $K=20$ variables for each asset, thus differing in the overall granularity of potential investments.
    \label{fig:asset_weights}
    }
\end{figure}

\begin{figure}
    \centering
    \includegraphics[width=\textwidth]{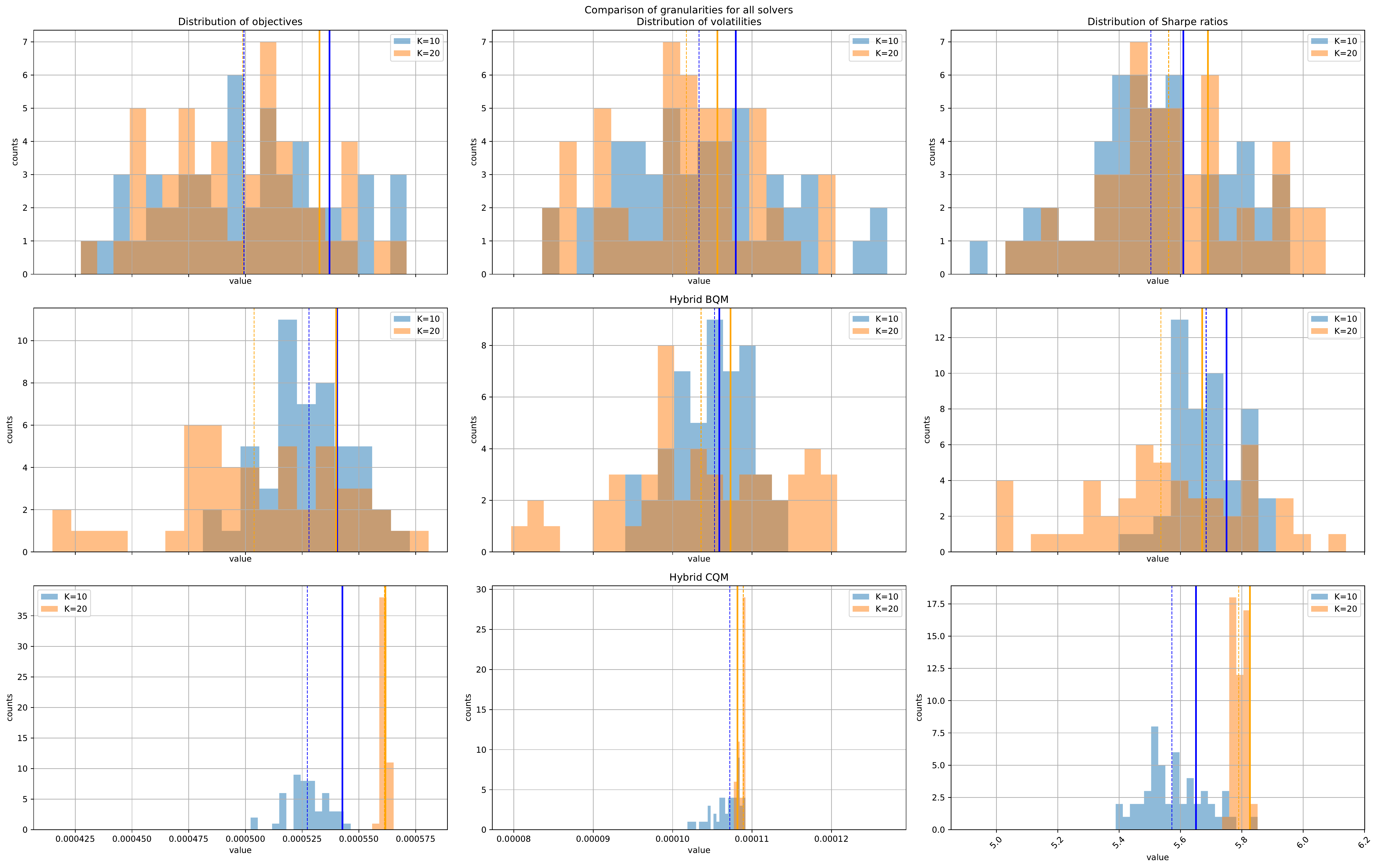}
    \caption{Comparison of objective (left column), volatility (middle column) and Sharpe ratios (right column) for solutions obtained with QBSolv (top row), Hybrid BQM (middle row) and Hybrid CQM (bottom row). Results for granularities $K=10$ (blue) and $K=20$ (orange) are reported. In each plot, the median value of the sample of solutions obtained is marked with a dashed line. The best acceptable portfolio from the sample, i.e. the portfolio with the highest objective obeying the volatility bound is marked with a solid line.}
    \label{fig:solvers_granularity}
\end{figure}

Given a fixed number of assets, employing more variables allows to increase the granularity of the weights of the individual assets, which should we expect to help satisfying the constraint that the sum of investments should equal $1$. This behaviour is confirmed by Figure~\ref{fig:asset_weights}, where we show the distributions of the sum of asset weights, i.e. the total amount of investment, both for $K=10$ and $K=20$ variables used to represent each asset. The peaks in the distribution for $K=20$ suggests that, for multiple solutions, all the solvers are able to find portfolios in which the total investment is closer to the constraint target value $1$. The statistics of the violation of the normalisation constraint for the three solvers and for two granularities are reported in  Table~\ref{tab:normviolation} along with our explicit calculation of the expected error in Appendix~\ref{sec:constraintviolation}. 

\begin{table}
\renewcommand{\arraystretch}{1.2}
\centering
\begin{tabular}{c | c c c c}
$1 - \textbf E(\sum w)$ & QBSolv & Hybrid BQM & Hybrid CQM & Theory\\
\hline
$K=10$ & $-4.89\cdot 10^{-5}$ & $-4.97\cdot 10^{-5}$ & $7.28\cdot 10^{-8}$ & $4.77 \cdot 10^{-7}$ \\
$K=20$ & $-7.20\cdot 10^{-9}$ & $-1.06\cdot 10^{-7}$ & $3.72\cdot 10^{-8}$ & $4.55 \cdot 10^{-13}$ \\
\hline
Variance \\
\hline
$K=10$ & $3.67 \cdot 10^{-9}$ & $1.42 \cdot 10^{-9}$ & $3.46 \cdot 10^{-13}$ & $3.97 \cdot 10^{-7}$ \\
$K=20$ & $4.48 \cdot 10^{-14}$ & $6.67 \cdot 10^{-15}$ & $1.67 \cdot 10^{-13}$ & $7.58 \cdot 10^{-13}$
\end{tabular}
\caption{Comparison of the violation of the normalisation constraint violation (deviation of mean of sum of weights from unity) and of the variance of the error for all the three solvers and for 10 and 20 binary variables, respectively. The results from an explicit calculation of the error expected due to finite granularity are reported as ``Theory''.}
\label{tab:normviolation}
\end{table}

Furthermore, while all solvers are able to find solutions having a relatively small deviation from the constraint target, QBSolv outputs solutions where such deviation is in the order of $10^{-4}$, BQM in the order of $10^{-5}$ and CQM, as the best solver, in the order of $10^{-6}$.

Comparing with the values measured from the distributions in Figure~\ref{fig:asset_weights} and reported in  Table~\ref{tab:normviolation}, we see that only a fraction of the error can be explained by the finite granularity.

\subsection{Observations}

Building on the study of the effect of the number of variables considered for each asset, Figure~\ref{fig:solvers_granularity} reports the results in terms of the KPIs for $K=10$ and $K=20$.
The first finding consists of the CQM solver providing more peaked distributions over multiple solutions when compared to other solvers, thus suggesting consistent - and high quality as can be seen from the measured values - results.
Then, we show that QBSolv and BQM do not report substantial differences in the distributions when compared to one another, while slight disparity is shown when comparing the distributions for $K=10$ and $K=20$ for each solver.

\section{Conclusions and Outlook}

In this work we have analyzed the capabilities of current Quantum and Hybrid solvers in solving the Portfolio Optimization problem. The data used represents a production environment and consists of 10 assets that can be divided in 3 main classes: equity, fixed-income and money market. This is a particularly interesting problem both in terms of common applicability in financial services as well as due to its nonlinear nature, which makes the QUBO formulation a particularly suitable model for the problem. We have thus detailed both the classical mathematical formulation of the Portfolio Optimization problem as well as the QUBO one.

We have explored the D-Wave's libraries and tackled the problem using the QBSolv, the Hybrid BQM and the Hybrid CQM solvers, while benchmarking the solutions with one given by exact classical methods. We have found that the CQM solver and its automating handling of multiple optimization terms and contraints QUBO can lead to higher quality solutions. Our satisfactory results show that the Quantum Computing approach is able to find solutions that are close to the exact optimum in terms of return and volatility.

These results pave the way for a broader applicability of the QUBO model using larger data sets, where a dramatic increase in the number of assets can lead classical solvers to yield only suboptimal solutions, while Quantum Computing is set to aim for high performing scaling capabilities and may thus outperform classical solutions in much more computationally-complex scenarios.
Concretely, next steps will include increasing the number of assets and the complexity of the problem. 







\bmhead{Acknowledgments}



The data and classical results were kindly provided by Simon Haller and Björn Chyba from RBI Research. We are also grateful for guidance on the business context and application. 
We also thank Vjekoslav Bonic for fruitful discussions.










\bibliography{sn-bibliography}

\newpage

\begin{appendices}

\section{Solution quality with respect to various parameters}

In this section, we examine how various model hyperparameters influence the solution quality.

First of all, the discrete approximation of the continuous weights can be expected to influence the solution quality.
In Figure~\ref{fig:solvers_granularity}, we report our findings on two choices of the granularity $K$. 
It appears that QBSolv does not profit a lot from increased granularity as the spread of the solution histograms is similar. However, the Sharpe ratio of the best usable portfolio is better for $K=20$, given the increased volatility for coarser granularity.

The picture is similar for Hybrid BQM.
The objectives are almost the same. Surprisingly, though, the Sharpe ratios are slightly better for lower granularity. Potentially, while using a larger granularity should in principle yield a better solution, it appears that the solver has troubles realizing this improvement as the solution space increases.

\begin{figure}
    \centering
    \includegraphics[width=\textwidth]{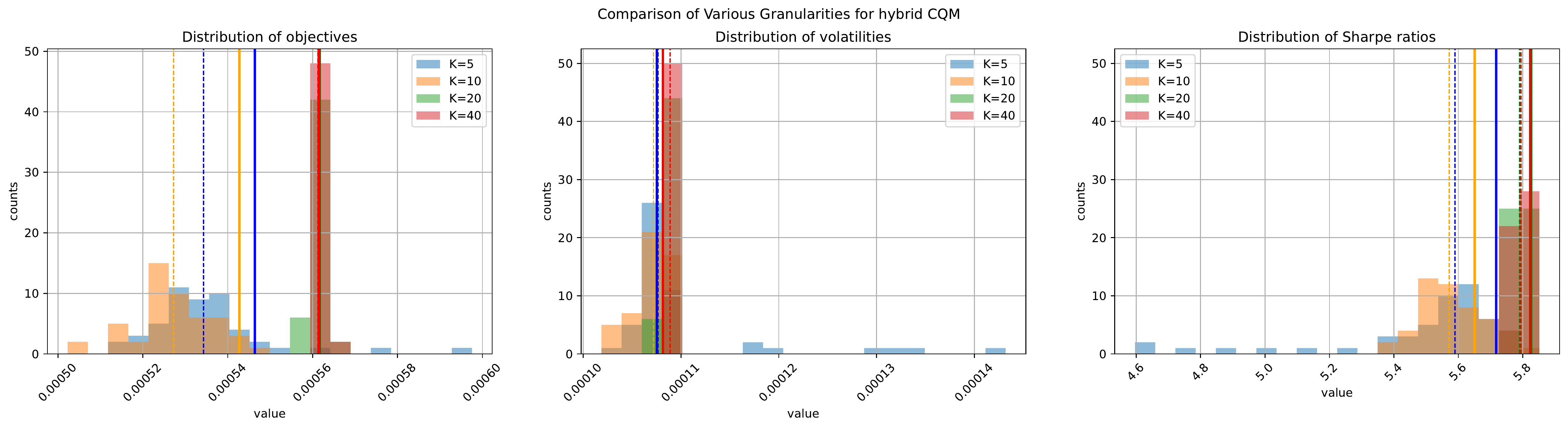}
    \caption{The effect of varying the granularity of the weights approximation for CQM. In general, higher granularities improve sampling of the solutions.}
    \label{fig:cqm_granularities}
\end{figure}

For Hybrid CQM, the effect is more pronounced. 
In general with increasing $K$ the solution gets better. The quality seems to saturate at $K=20$. It seems an accidental finding that $K=5$ outperforms $K=10$. It is clearly visible that for $K=5$ the spread of the solutions is very large as can be seen in Figure~\ref{fig:cqm_granularities}.

\section{Iterations}

\begin{figure}
    \centering
    \includegraphics[width=\textwidth]{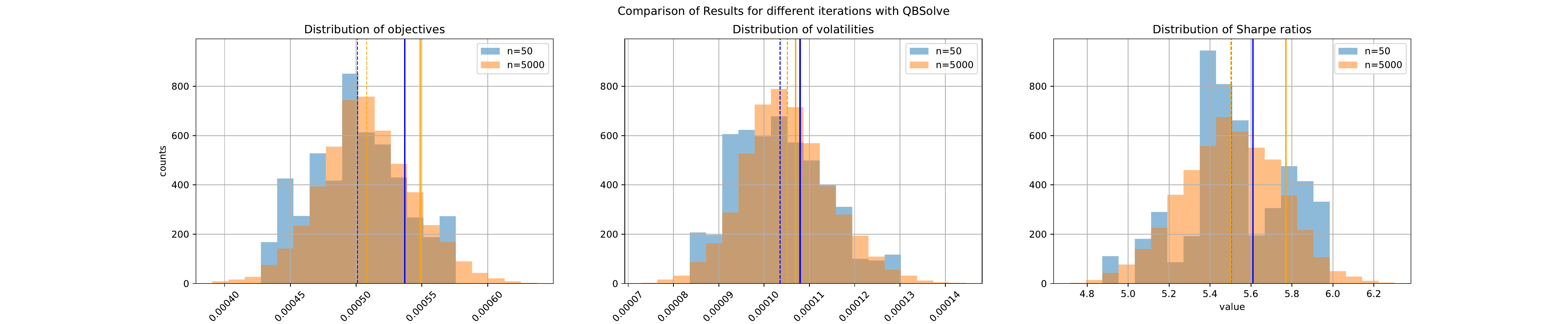}
    \caption{Comparison of solution quality after increasing the number of iterations for QBSolv. A slightly better solution can be found by sampling more often. Histograms are scaled for comparability.}
    \label{fig:iterations}
\end{figure}

\begin{figure}
    \centering
    \includegraphics[width=0.48\textwidth]{figures/risk_vs_return/vola_vs_return_QBSolv_v03_a010_w10_n50_t05.pdf}
    \includegraphics[width=0.48\textwidth]{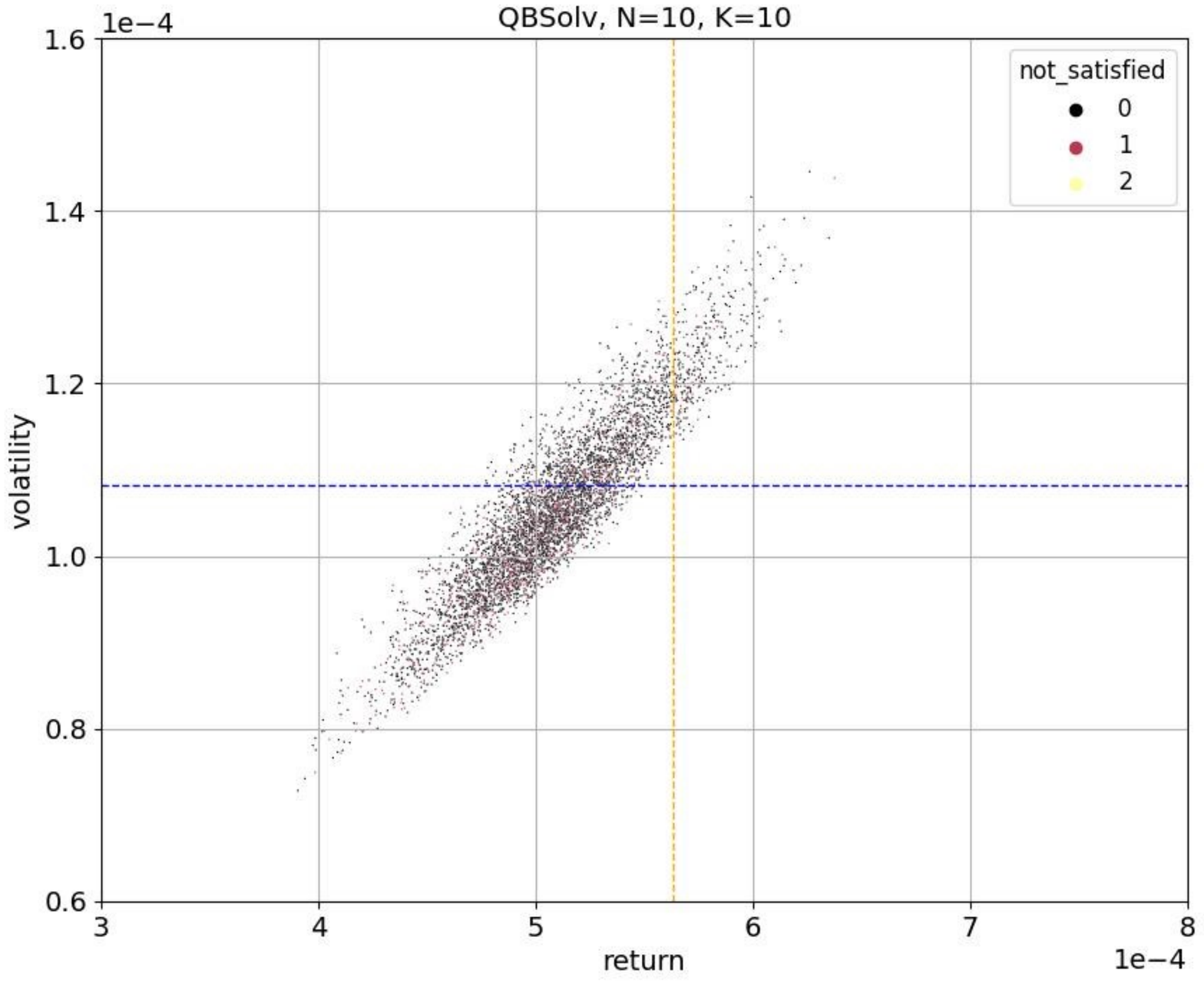}
    \caption{Comparing the solutions for QBSolv in risk vs. volatility for 50 (left) and 5000 (right) iterations, respectively. It can be seen qualitatively that a lower sampling provides already a good representation of the solution space.}
    \label{fig:scatter_iterations_qbsolv}
\end{figure}

It becomes clear from Figure~\ref{fig:iterations} that sampling the result for 50 iterations is not too far off a better result obtained with 5000 iterations at correspondingly higher costs. Figure \ref{fig:scatter_iterations_qbsolv} also shows this qualitatively. Therefore it seems that a smaller sampling is already sufficient.

\section{Constraint Violation}
\label{sec:constraintviolation}

In Section~\ref{sec:problemformulation}, we have presented the constraints on the problem solution. Those constraints are hard constraints for the business context and a solution can only be used if they are obeyed. Due to the nature of the solution strategy, however, violation of some of the constraints is to be expected. This is because the constraints are included in QUBO by imposing an energy penalty, which does not guarantee it is obeyed. Therefore, post selection of the results is necessary. For the application in a production context, it is relevant to understand the success probability in the sense of which fraction of the results are not violating any constraints.

\begin{figure}
    \centering
    \includegraphics[width=0.8\textwidth]{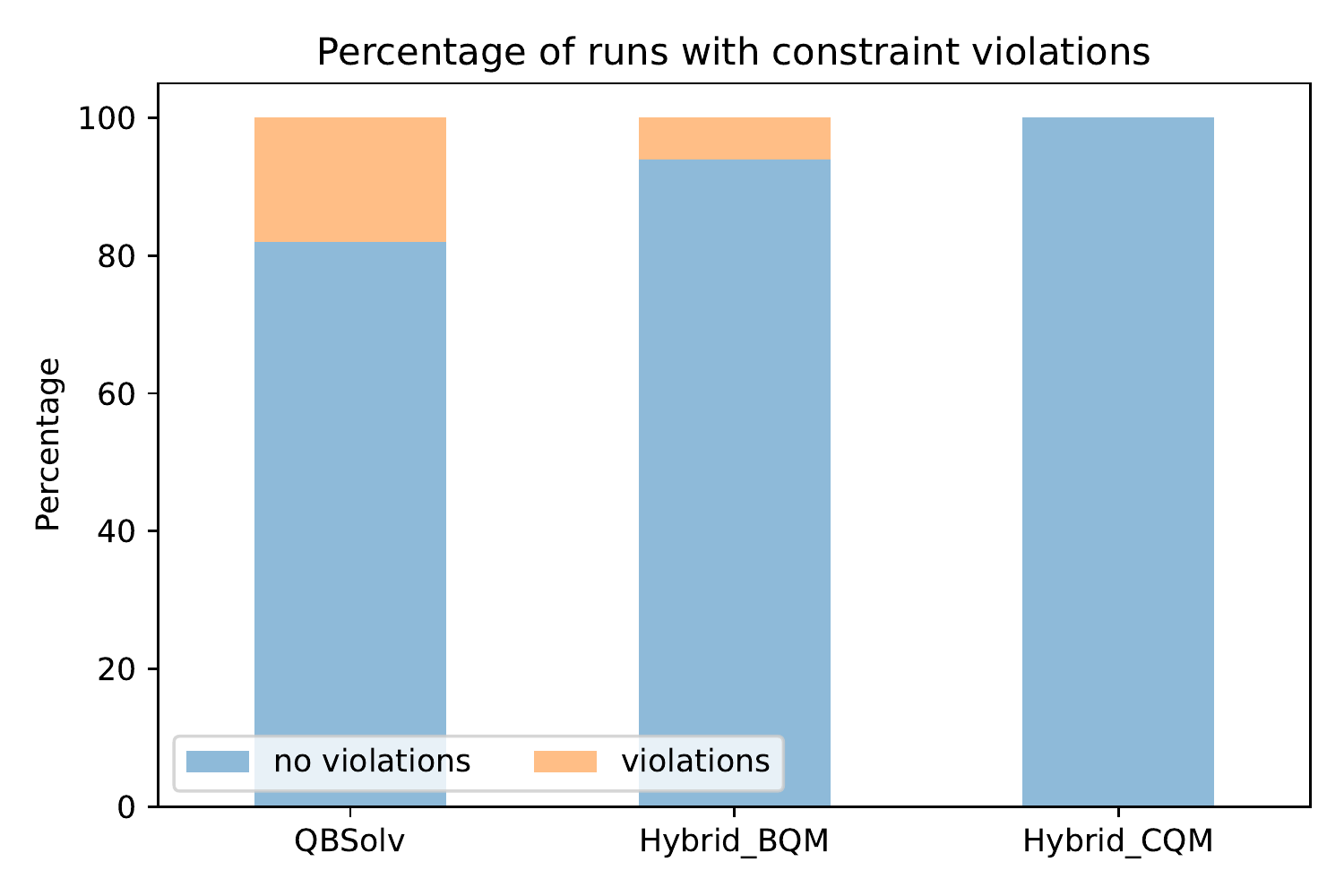}
    \caption{Comparison of success probability for each solver. Any constraint violation will be counted. While in QBSolv and Hybrid BQM, 82\% and 96\% of the runs were satisying all the constrains, respectively, this was true for all runs with Hybrid CQM.}
    \label{fig:violation_all_solvers}
\end{figure}

We are evaluating the constraint violations found in our experiments in Figure~\ref{fig:violation_all_solvers}. With success probabilities between 82 and 100 percent, the procedure is usable with large enough sample size. The hybrid methods have a slightly higher success probability than the simulation, which manifests the utility of using the QPU in the calculation. Hybrid CQM (100\%) has a slight advantage over Hybrid BQM (94\%). Here, we have evaluated only runs with 10 assets and 10 qubits.

\begin{figure}
    \centering
    \includegraphics[width=0.8\textwidth]{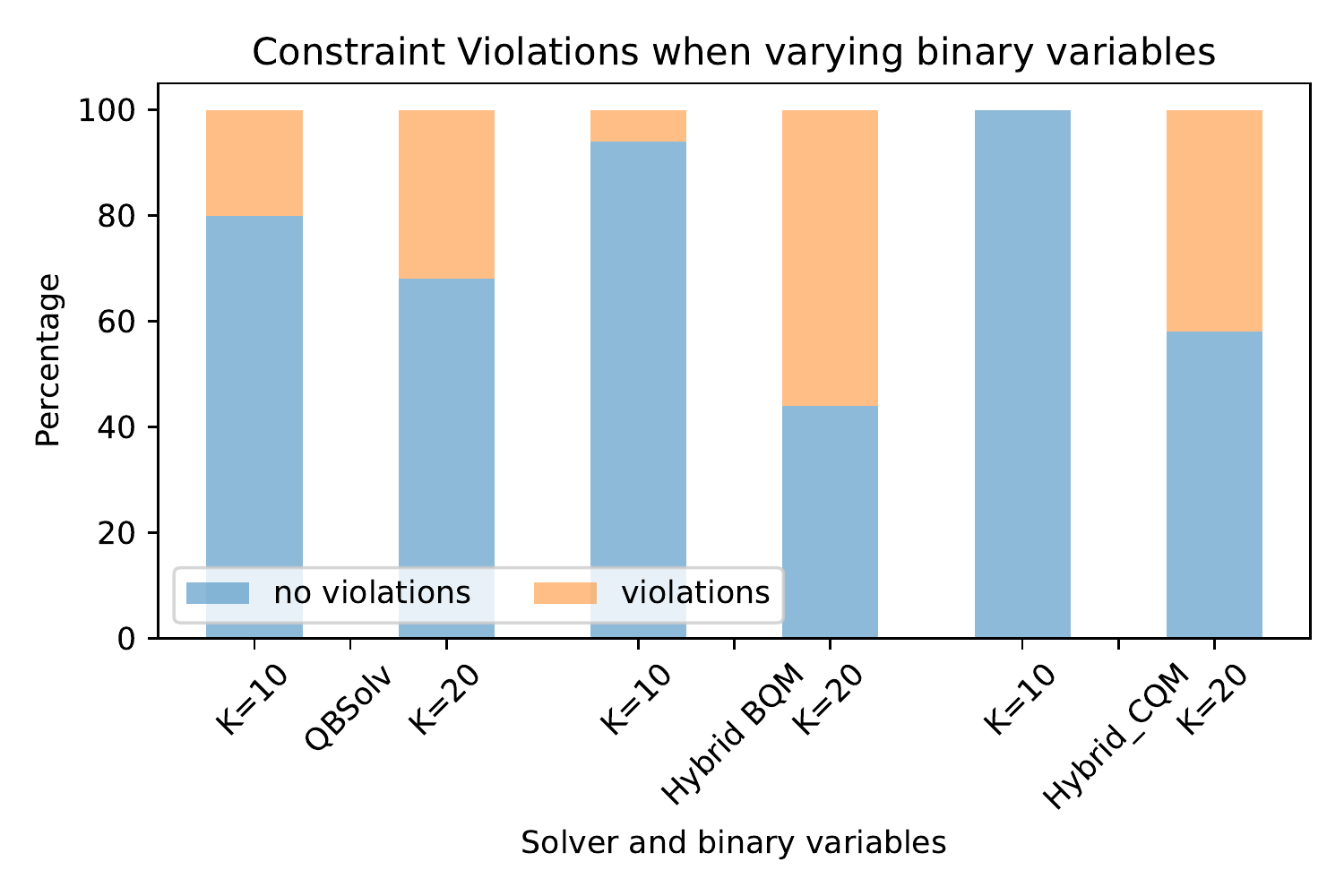}
    \caption{Comparison of success probability for different numbers of binary variables for each of the solvers. It is striking to see that adding more variables is not necessarily leading to better performance. Presumably, this is due to the increase in search space, which makes it more difficult for the solver to find a solution satisfying all constraints.}
    \label{fig:violation_w}
\end{figure}

\begin{figure}
    \centering
    \includegraphics[width=0.8\textwidth]{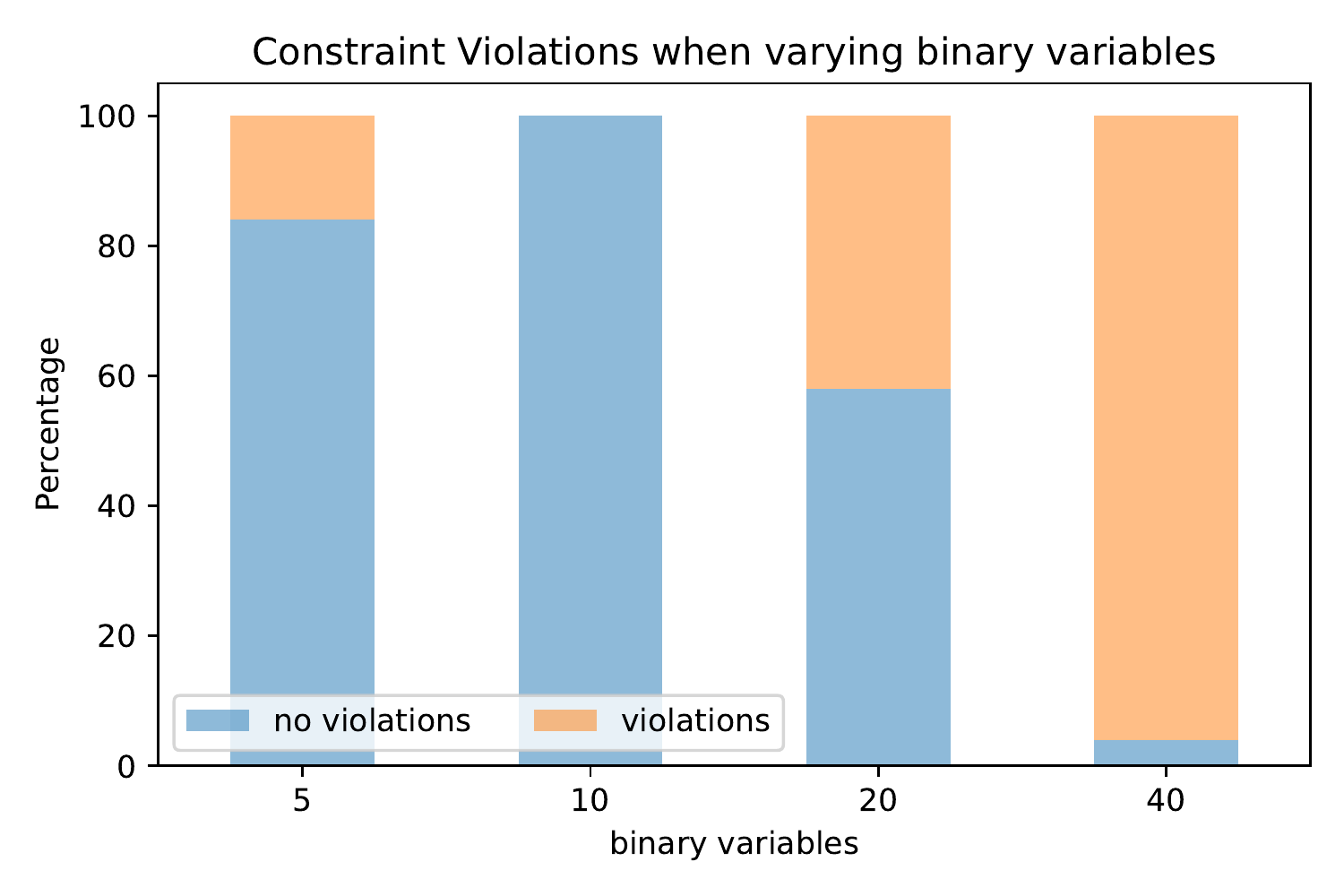}
    \caption{The effect of the granularity on constraint violation for Hybrid CQM.}
    \label{fig:violation_CQM_w}
\end{figure}

Including also higher number of assets and binary variables shows a much more diversified picture. In Figure \ref{fig:violation_w} we make the rather unexpected observation that satisfying all constraints becomes more difficult with more binary variables. For Hybrid CQM, we have examined even more values for $K$ in Figure \ref{fig:violation_CQM_w}. 

\begin{figure}
    \centering
    \includegraphics[width=0.8\textwidth]{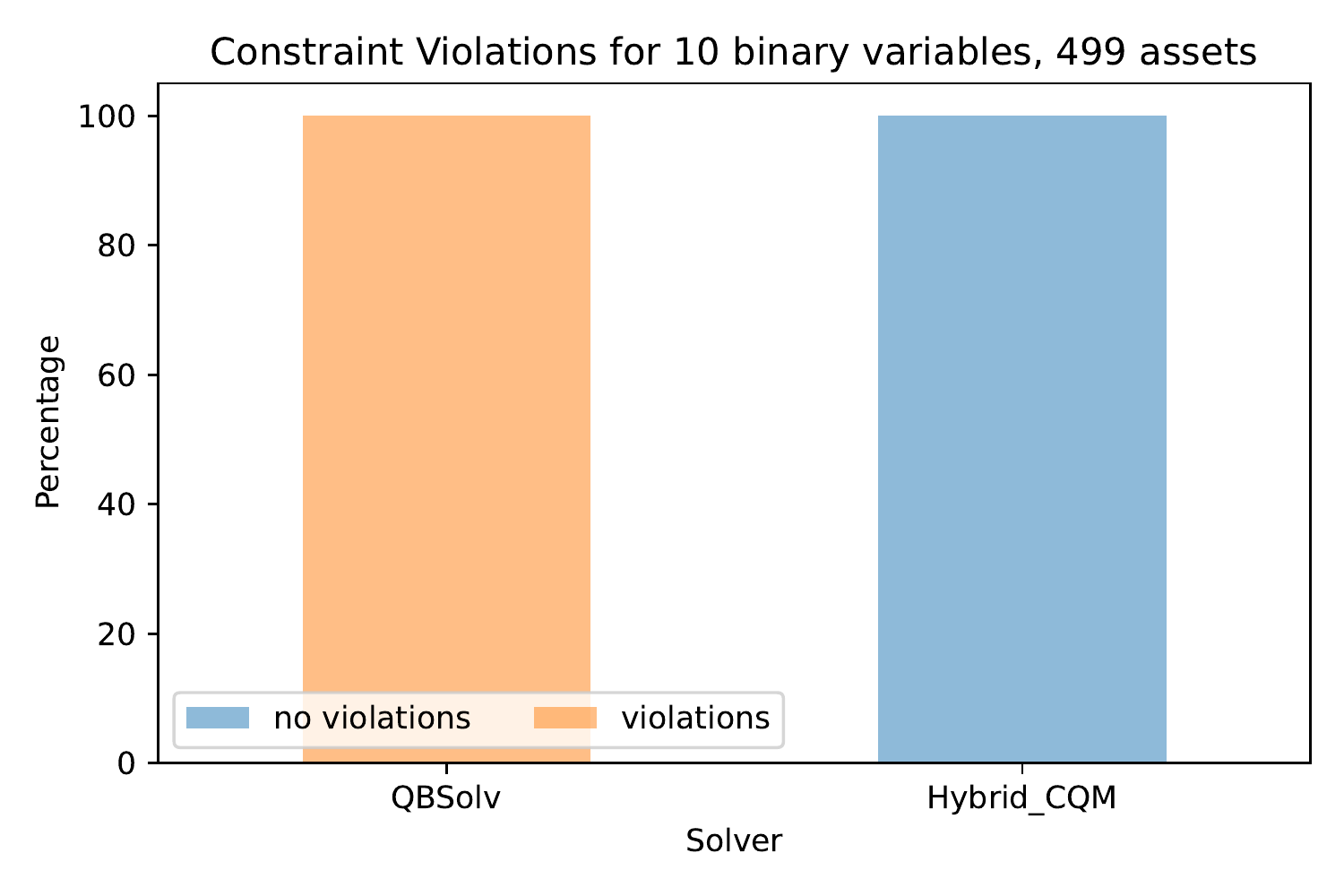}
    \caption{Examination of the success probability for a large number of assets ($N=499$) shows that Hybrid CQM can still find solutions which satisfy the constraints as opposed to QBSolv.}
    \label{fig:violation_a499}
\end{figure}

An interesting experiment is to scale the number of assets, because we expect the quantum computer to be more performant than the classical solver when we increase this number. In Figure~\ref{fig:violation_a499}, we see that QBSolv is not able to find any permissible solutions with 499 assets while Hybrid CQM still always finds a permissible solution.

\begin{figure}
    \centering
    \includegraphics[width=0.8\textwidth]{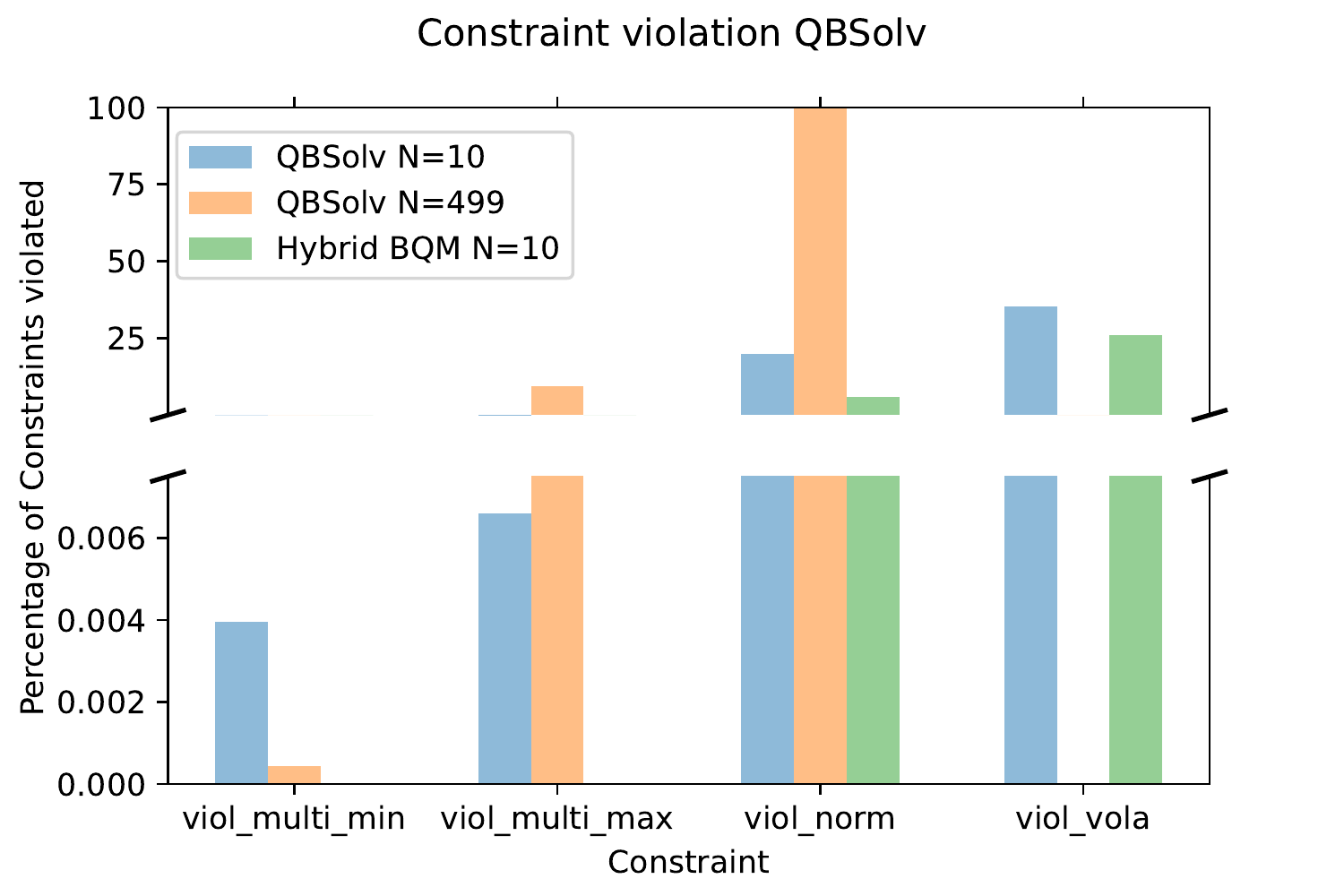}
    \caption{Percentage of experiments with constraint violations per constraint type and solver. Multi-min and multi-max constraints are violated at a much lower frequency than normalisation and volatility constraint.}
    \label{fig:violation_constraints}
\end{figure}

For the application of the procedure in a business context, not all violations are equally problematic. The volatility and normalisation constraints can be slightly violated, for instance, while regulatory constraints must not. We therefore also examine which constraints are violated in each context in Figure~\ref{fig:violation_constraints}. 

Given the construction in \eqref{eq:disc_weight}, the single-min and single-max constraints are automatically and always satisfied and are therefore not displayed. This is not true for the other constraints. We see violations on two levels. While the multi-min and multi-max constraints are violated on the $10^{-3}$\% level (except for QBSolv with many assets), normalisation and volatility constraints are violated above 20\%. 

The normalisation constraint is affected by the granularity as we have already detailed in Subsection~\ref{sec:disc}. We report the violation of the normalisation constraint in our experiments in Figure~\ref{fig:asset_weights}.

\begin{figure}
    \centering
    \includegraphics[width=0.8\textwidth]{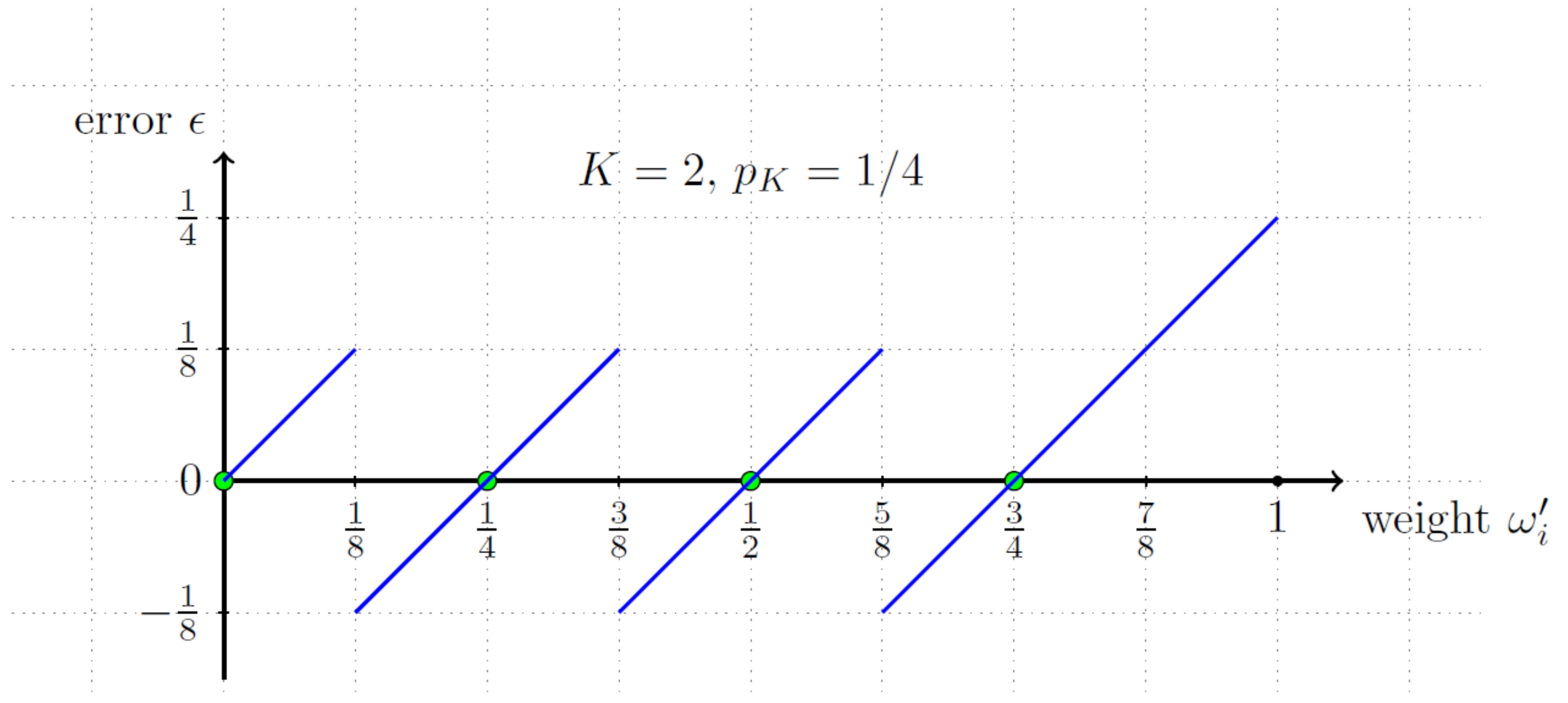}
    \caption{This sketch explains the calculation of the error in the approximation of a rational number between $0$ and $1$ using binary variables. For the purpose of the illustration we are using $K=2$ variables without loss of generality. Potential representations $(0,0), (1,0), (0,1), (1,1)$ which represent the numbers $0, \frac{1}{4}, \frac{1}{2}, \frac{3}{4}$ (green dots). Each possible continuous rational number is represented by a discrete binary value that has the smallest difference. The difference/error behaves as depicted and grows larger close to unity. This is because in the construction of Section~\ref{sec:disc} unity is not reached due to a trade-off for higher resolution.}
    \label{fig:telescope}
\end{figure}

It is obvious that the granularity affects the violation of the normalisation constraint. For the calculation of the error stemming from the finite granularity of representing the binary expansion of the weights, we are applying a telescope procedure explained in Figure~\ref{fig:telescope}. 

In principle, every rational number between 0 and 1 is equally probable for a specific weight, so the probability distribution is uniform. The error produced when representing the continuous number by a discrete binary variable depends in a linear fashion on its distance from the number. The representation in Figure~\ref{fig:telescope} makes it directly obvious, that the expected error cancels in every part of the telescope.

%
%
%
%

We are first calculating the expected value of the error $\epsilon$ as depicted in Figure~\ref{fig:telescope}. 
Each contribution to the expectation value $\E[\epsilon]$ from the first $2^K-1$ integrals cancel. We see that
\begin{align}
    \E_0 &= \int_0^{\frac{p_K}{2}} x \, \de x + \int_{\frac{p_K}{2}}^{p_K} (x-p_K) \, \de x \nonumber \\
    &= 0 \nonumber \\
    &= \E_i \quad \text{for} \; i\leq2^K-1 \, .
\end{align}
The non-zero contribution to the expected value is
\begin{equation}
\label{avg_error}
    \E_{2^K} = \E[\epsilon] = \int_{1-p_K}^1 (x-(1-p_K)) \, \de x = \frac{p_K^2}2 \, \sim \, \mathcal{O}(p_K^2) \, .
\end{equation}
Note that this is the expected error for an individual unconstrained weight with possible values in $[0,1]$. We obtain the expected value of the sum of weights according to \eqref{eq:minmaxerror}. In the case at hand in this study, it turns out that the expected error of the normalisation constraint is the same as the pre-factor just turns out to be unity.

Concerning the standard deviation, we are determining the second moment of our error distribution $\Var[\epsilon] = \E[\epsilon^2] - \E[\epsilon]^2$. We are following the prescription outlined in Figure~\ref{fig:telescope}. The first integral of $\E[\epsilon^2]$
\begin{align}
    \E[\epsilon^2]_0 &= \int_0^{\frac{p_K}2} x^2 \, \de x + \int_{\frac{p_K}2}^{p_K} (x - p_K)^2 \, \de x = \frac{p_k^3}{12} \nonumber \\
    &= \E[\epsilon^2]_i \quad \text{for}\; i\leq2^k-1 \, .
    \label{eq:variance_e20}
\end{align}
Given the structure of the telescope, this is the contribution to the variance for all the integrals up to the final one
\begin{equation}
   \E[\epsilon^2]_{2^K} = \int_{1 - p_K}^1 (x - (1 - p_K))^2 \, \de x =  \frac{p_K^3}3 \, .
   \label{eq:variance_e22k}
\end{equation}
Summing $2^K-1$ parts of \eqref{eq:variance_e20} and \eqref{eq:variance_e22k} we obtain
\begin{align}
   \E[\epsilon^2] & =  \left(\frac1{p_K}-1 \right) \frac{ p_K^3}{12}  + \frac{p_K^3}3 \nonumber \\
   & = \frac{p_K^2}{12} + \frac{p_K^3}4 \, \sim \, \mathcal{O}(p_K^2) \, .
\end{align}

Putting it all together, we obtain for the variance for the error
\begin{equation}
\label{var_error}
    \Var[\epsilon] = \frac{p_K^2}{12} + \frac{p_K^3}4 - \frac{p_K^4}{4} \, \sim \, \mathcal{O}(p_K^2) \, .
\end{equation}
For determining the skewness
\begin{equation}
\label{skew}
\Skew[\epsilon] = \frac{\E[\epsilon^3] - 3\E[\epsilon] \Var[\epsilon] - \E[\epsilon]^3}{(\Var[\epsilon])^{3/2}} \, ,
\end{equation}
we calculate the third moment $\E[\epsilon^3]$. Due to the anti-symmetry of the error function, the first integrals vanish when partitioning in the same way as when calculating the expectation value
\begin{align}
 \E[\epsilon^3]_i &= \int_0^{\frac{p_K}2} x^3 \, \de x + \int_{\frac{p_K}2}^{p_K} (x - p_K)^3 \, \de x = 0 \quad \text{for}\; i \leq 2^K-1 \;. 
\end{align}
The non-vanishing contribution comes from
\begin{align}
\label{third_moment_error}
 \E[\epsilon^3]_{2^K} &= \int_{1 - p_K}^1 (x - (1 - p_K))^3 \, \de x = \frac{p_K^4}4 \, .
\end{align}
Inserting the results of (\ref{avg_error}), (\ref{var_error}), and (\ref{third_moment_error}) into (\ref{skew}) one gets
\begin{eqnarray}
\Skew[\epsilon] &=& \frac{\frac{p_K^4}{4} - 3 \frac{p_K^2}{2} (\frac{p_K^2}{12} + \frac{p_K^3}{4} - \frac{p_K^4}{4}) - (\frac{p_K^2}{2})^3}{(\frac{p_K^2}{12} + \frac{p_K^3}{4} - \frac{p_K^4}{4})^{\frac{3}{2}}} = \nonumber \\
&=& 3 \sqrt{3} \, (p_K - 3 p_K^2 + 2 p_K^3) \cdot (1 + 3p_K - 3p_K^2)^{-\frac{3}{2}} \nonumber \\ 
&=& 3 \sqrt{3} \, p_K + \mathcal{O}(p_K^2) \, \sim \, \mathcal{O}(p_K) \, .
\end{eqnarray}

All in all, the considerations in this shows good usability of the approach for a large enough sampling, which is reflected already in Figure~\ref{fig:solvers_granularity}, where the best usable portfolio, which is marked with a solid line is chosen such that no constraints are violated.


\section{Zoomed Plots}

\begin{figure}
    \centering
    \includegraphics[width=0.6\textwidth]{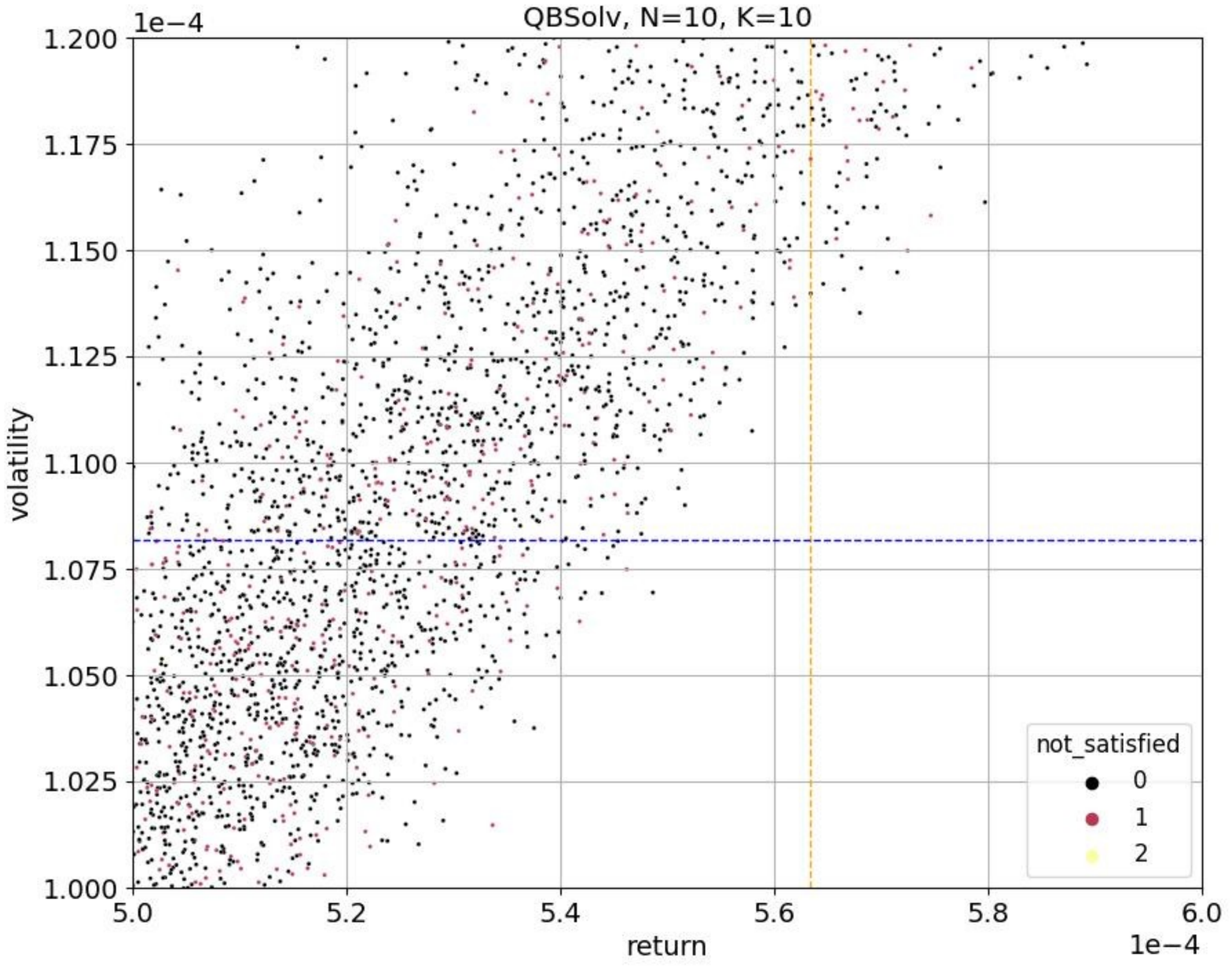}
    \includegraphics[width=0.6\textwidth]{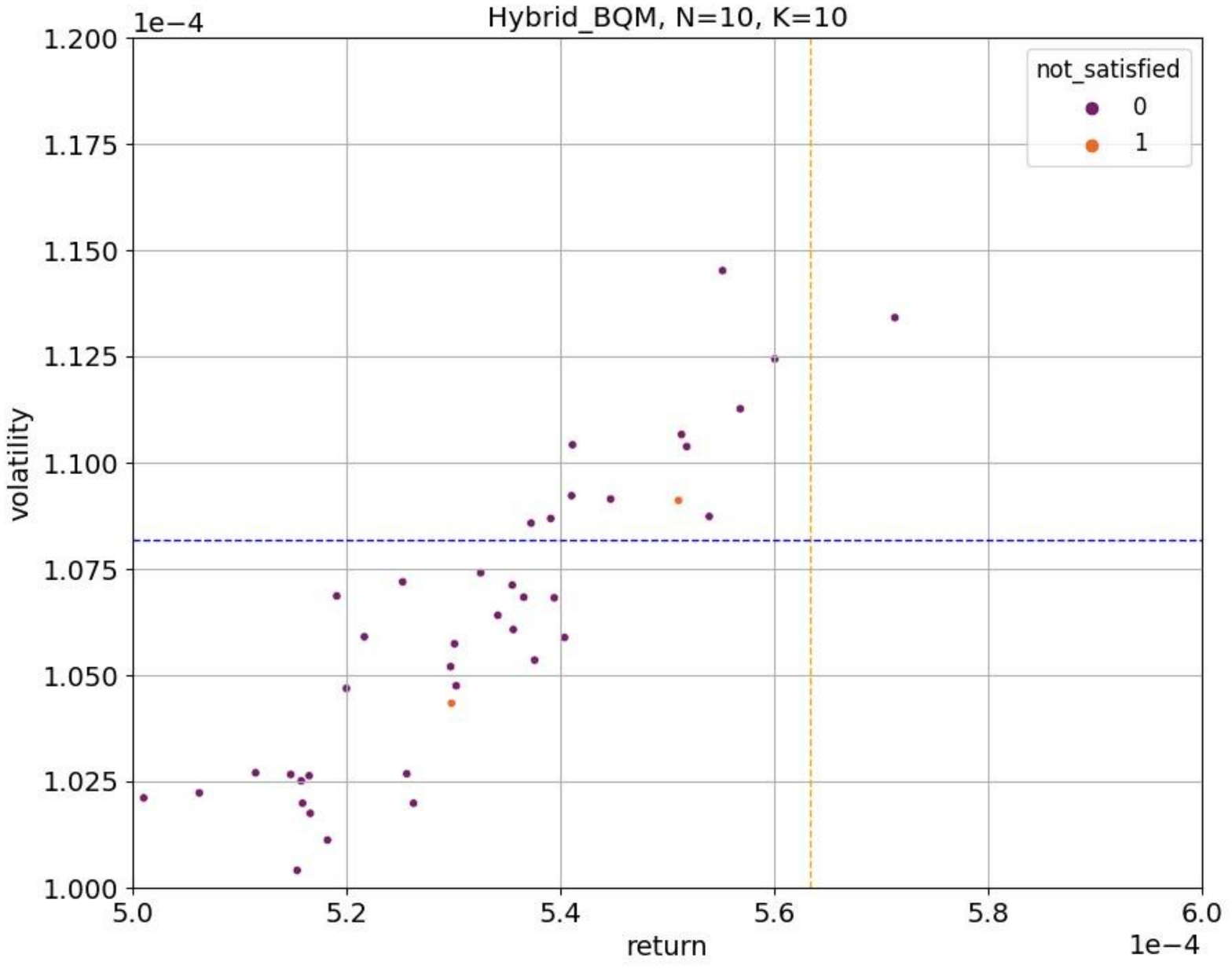}
    \includegraphics[width=0.6\textwidth]{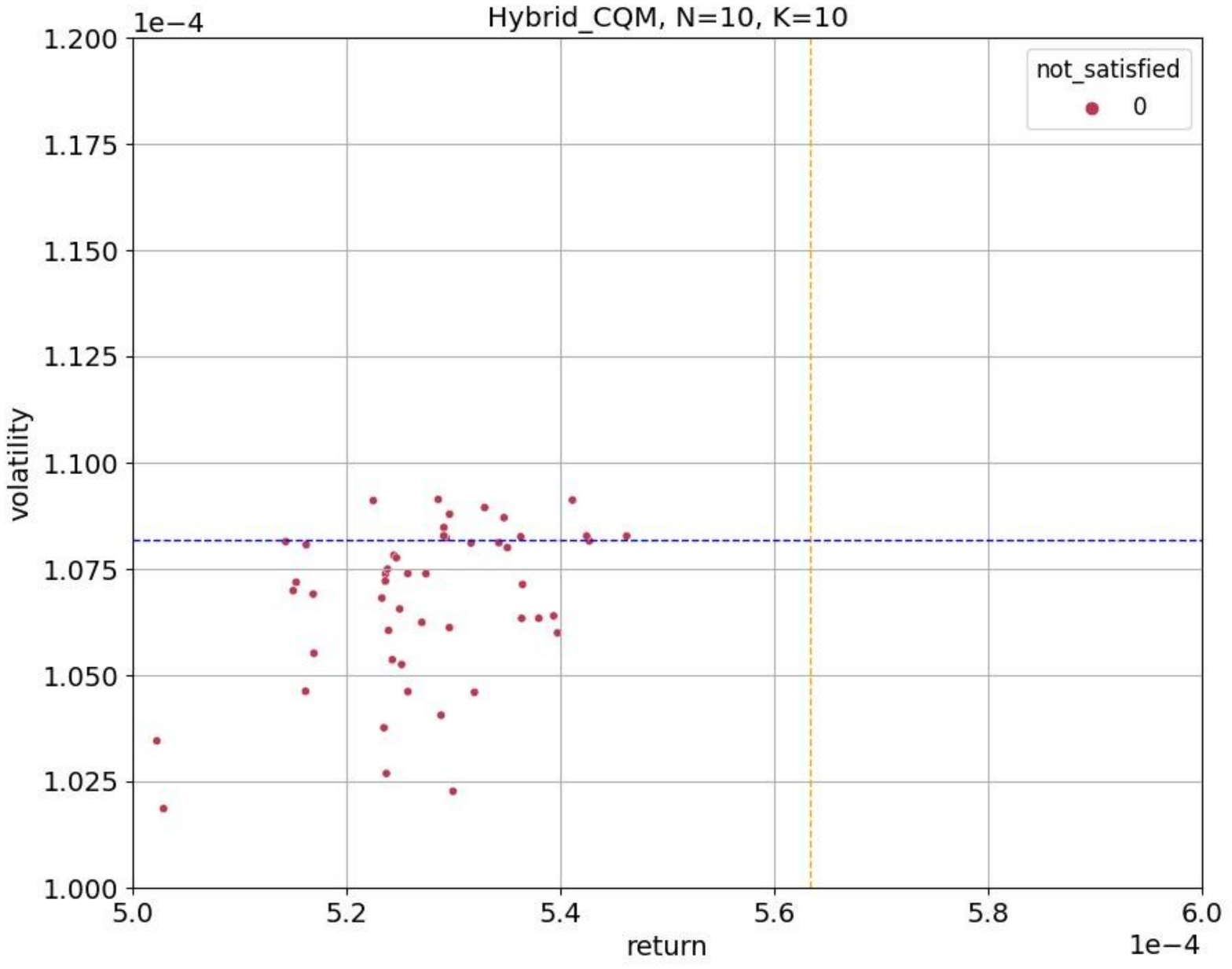}
    \caption{Return is plotted against volatility like in Figure \ref{fig:result_scatterplot} but zoomed in around the center of the solutions for better visibility. It can be seen very clearly that the variance of the solutions is greatly reduced by the use of a QPU. In particular for Hybrid CQM, the solutions become more clustered around the volatility threshold line and the shape is more circular than elongated, which gives a hint at the different way the quantum algorithm is used.}
    \label{fig:result_scatter_zoomed}
\end{figure}

%
%
%

\end{appendices}
\end{document}